%% file: mulsk.tex
\documentclass[aps,prb,twocolumn,groupedaddress,amsmath,amssymb, superscriptaddress]{revtex4-1}

\usepackage{graphicx}
\usepackage{amsmath}
\usepackage{amssymb}
\usepackage{bm}% bold math
\usepackage{dcolumn}% Align table columns on decimal point
\usepackage{soul} % strikethough, highlight
\usepackage{psfrag}
\usepackage{float}
\usepackage{subeqnarray}
\usepackage{url}
\usepackage[colorlinks=true,linkcolor=blue,citecolor=blue,urlcolor=blue,breaklinks]{hyperref}

\begin{document}

%\preprint{}

\title{Paths to collapse for isolated skyrmions in few-monolayer ferromagnetic films}

\author{Dusan Stosic}
\affiliation{Centro de Inform\'atica, Universidade Federal de Pernambuco, Av. Luiz Freire s/n, 50670-901, Recife, PE, Brazil}
\affiliation{Departement Fysica, Universiteit Antwerpen, Groenenborgerlaan 171, B-2020 Antwerpen, Belgium}
\author{Jeroen Mulkers}
\affiliation{Departement Fysica, Universiteit Antwerpen, Groenenborgerlaan 171, B-2020 Antwerpen, Belgium}
\affiliation{DyNaMat Lab, Department of Solid State Sciences, Ghent University, Ghent, Belgium}
\author{Bartel Van Waeyenberge}
\affiliation{DyNaMat Lab, Department of Solid State Sciences, Ghent University, Ghent, Belgium}
\author{Teresa Ludermir}
\affiliation{Centro de Inform\'atica, Universidade Federal de Pernambuco, Av. Luiz Freire s/n, 50670-901, Recife, PE, Brazil}
\author{Milorad V. Milo\v{s}evi\'c}
\email[Email: ]{milorad.milosevic@uantwerpen.be}
\affiliation{Departement Fysica, Universiteit Antwerpen, Groenenborgerlaan 171, B-2020 Antwerpen, Belgium}


%\author{Dusan Stosic}
%\email[]{db.stosic@gmail.com}
%\thanks{}
%\affiliation{}


\date{\today}

\begin{abstract}
Magnetic skyrmions are topological spin configurations in materials with chiral Dzyaloshinskii-Moriya interaction (DMI), that are potentially useful for storing or processing information. To date, DMI has been found in few bulk materials, but can also be induced in atomically thin magnetic films in contact with surfaces with large spin-orbit interactions. Recent experiments have reported that isolated magnetic skyrmions can be stabilized even near room temperature in few-atom thick magnetic layers sandwiched between materials that provide asymmetric spin-orbit coupling. Here we present the minimum-energy path analysis of three distinct mechanisms for the skyrmion collapse, based on ab initio input and the performed atomic-spin simulations. We focus on the stability of a skyrmion in three atomic layers of Co, either epitaxial on the Pt(111) surface, or within a hybrid multilayer where DMI nontrivially varies per monolayer due to competition between different symmetry-breaking from two sides of the Co film. In laterally finite systems, their constrained geometry causes poor thermal stability of the skyrmion toward collapse at the boundary, which we show to be resolved by designing the high-DMI structure within an extended film with lower or no DMI. 
\end{abstract}

\pacs{}
%\keywords{}

\maketitle

%\section{}
%\subsection{}
%\subsubsection{}

\section{Introduction\label{intro}}
In ferromagnetic materials with broken inversion symmetry, electron spins can twirl to form magnetic topological defects known as skyrmions~\cite{Rossler}. These novel objects were initially observed in bulk magnetic materials with non-centrosymmetric crystal lattices~\cite{Muhlbauer,Pappas,Moskvin}, but have more recently been found in ultrathin magnetic films on surfaces with strong spin-orbit interactions~\cite{Heinze,Romming}. Skyrmions are typically formed due to the competition of the exchange and the Dzyaloshinskii-Moriya (DM) interactions~\cite{Dzyaloshinsky,Moriya}. The small size (down to a few nanometers)~\cite{Wiesendanger} and the very small electrical currents~\cite{Jonietz,Iwasaki} needed to displace them, increase the appeal of skyrmions for prospective spintronic applications.

Magnetic skyrmions have also been proposed as information carriers in possible data-processing devices~\cite{Kiselev}. For example, in racetrack memories one can encode information in a magnetic track by a train of skyrmions which represent individual bits of information~\cite{Sampaio,Fert}. These skyrmions are then moved by an applied current and information is read from their electrical signatures~\cite{Nagaosa}. One may also be able to create skyrmion-based logic devices~\cite{Zhang} or magnetic memories that operate in a similar way to phase change memories~\cite{Liang1}. However, the realization of any practical applications will require the possibility to create and destroy individual skyrmions as well as to manipulate their positions. By tuning the relative strength of competing interactions active in the magnetic material, it has already been shown that individual nanoscale skyrmions can be stabilized~\cite{Sampaio,Romming1}. It is also possible to wire and delete such skyrmions~\cite{Romming} and to precisely control their motion using electrical currents~\cite{Jonietz} or local electric fields~\cite{Hanneken}.

The concept of topology provide useful insights into the stability of an isolated skyrmion. States that are termed topologically protected mean there is an energy barrier separating the transition of a system from one topological state to another. When considering the magnetization to be a continuous vector field, which is the basic assumption in the micromagnetic framework, it is impossible to collapse the skyrmion on itself in a continuous manner. At some point in the process, a Bloch point has to appear, resulting in an infinite exchange-energy contribution. For this reason, skyrmionic (Sk) states are believed to be topologically protected from other competing spin configurations such as ferromagnetic (FM) and spiral (helical or conical) orders~\cite{Oike}. In real materials, however, the system is discrete and magnetic moments are localized on atoms, resulting in finite energy barriers that can be surpassed by thermal fluctuations. It is important to understand these energy barriers in order to evaluate the stability and lifetime of isolated skyrmions, and possibilities for their further manipulation. Since the micromagnetic framework is not suited to model transitions accompanied with a Bloch point, we have to resort to more accurate, but computationally intensive, atomistic simulations to study the energy barriers between states with a different topology.

Single magnetic skyrmions have already been extensively studied both in theory~\cite{Sampaio,Hagemeister} and in experiments~\cite{Romming,Romming1}. Very recently Refs.~\cite{Rohart,Lobanov} reported detailed descriptions of the mechanisms for skyrmion collapse in a single atomic layer of magnetic material with interfacially-induced DM interaction~\cite{Dupe1}. These restricted conditions can severely limit the use of skyrmions in practical applications because (a) creating large-area monolayers requires specialized preparation which will make commercial production hard to achieve and (b) isolated skyrmions in magnetic monolayers have only been observed at very low temperatures. For realistic applications, room-temperature skyrmions can be achieved by increasing the thickness of the magnetic film to few monolayers and increasing the interfacial spin-orbit coupling by using two different materials below and above the film~\cite{Moreau,Boulle}. In view of the emergent importance of such multilayer hybrid structures, a deep investigation of the behavior of isolated magnetic skyrmions within few-atom thick ferromagnets with inhomogeneous DMI is crucial for the design of future spintronic devices.

In this paper we analyze the thermal stability of an isolated skyrmion in ultrathin chiral magnetic structures composed of few monolayers. We focus on the realistic system of three atomic layers of Co, grown epitaxially on the Pt(111) surface, with or without a capping material with associated spin-orbit coupling. Realistically, the DM interaction is diluted from one Co monolayer to another, correspondingly to related ab initio calculations in the literature~\cite{Yang,Boulle}. Using atomic scale spin simulations, we demonstrate that the properties of magnetic skyrmions, such as their diameter and their stability, are significantly different compared to the case of a Co monolayer (or constant DM interaction in a thicker film). To better understand the stability of skyrmions in few-atom thick films and hybrid multilayers, we study their collapse mechanisms by minimum-energy path calculations and reveal three distinct collapse paths (on itself, on the boundary, and at an interface). 

The paper is organized as follows. In Sec. \ref{model} we provide the brief description of our theoretical framework and numerical simulations. Sec. \ref{stab} is devoted to the analysis of skyrmion stability and size in a 3ML Co film, and its dependence on the vertical profile of the DM interaction, and exchange and anisotropy effects. In Sec. \ref{coll} we discuss the energy paths for skyrmion collapse (on itself, at a lateral sample boundary, or at a lateral interface where DMI vanishes) in the same parametric domain as in Sec. \ref{stab}, and discuss the thermal stability and lifetime of skyrmions in the different cases. Our findings are summarized in Sec. \ref{concl}.

\section{Model\label{model}}
\begin{figure}
\begin{center}
\includegraphics[width=0.16\columnwidth]{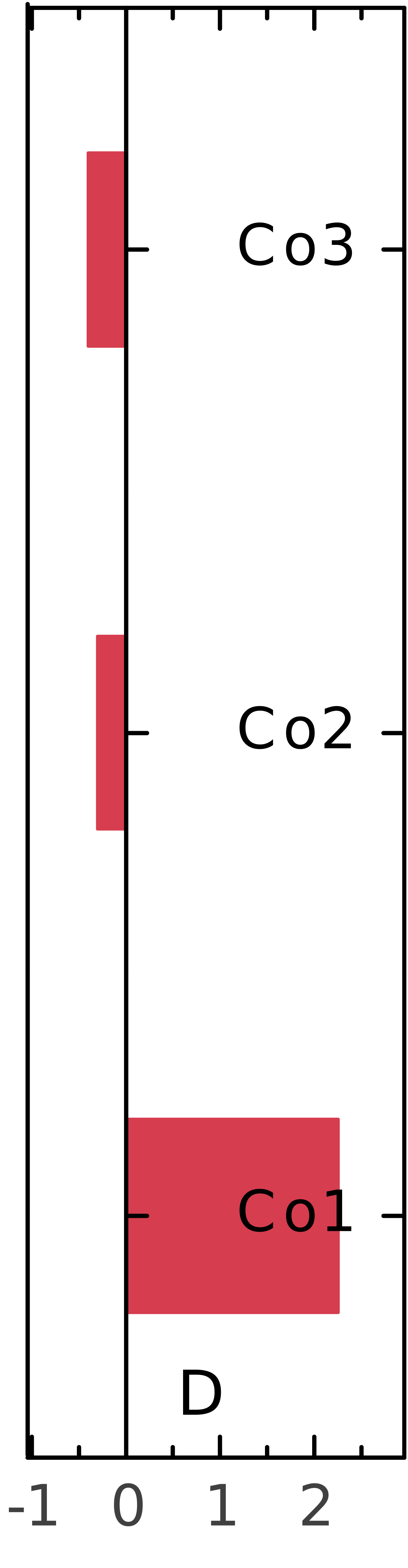}
\includegraphics[width=0.82\columnwidth]{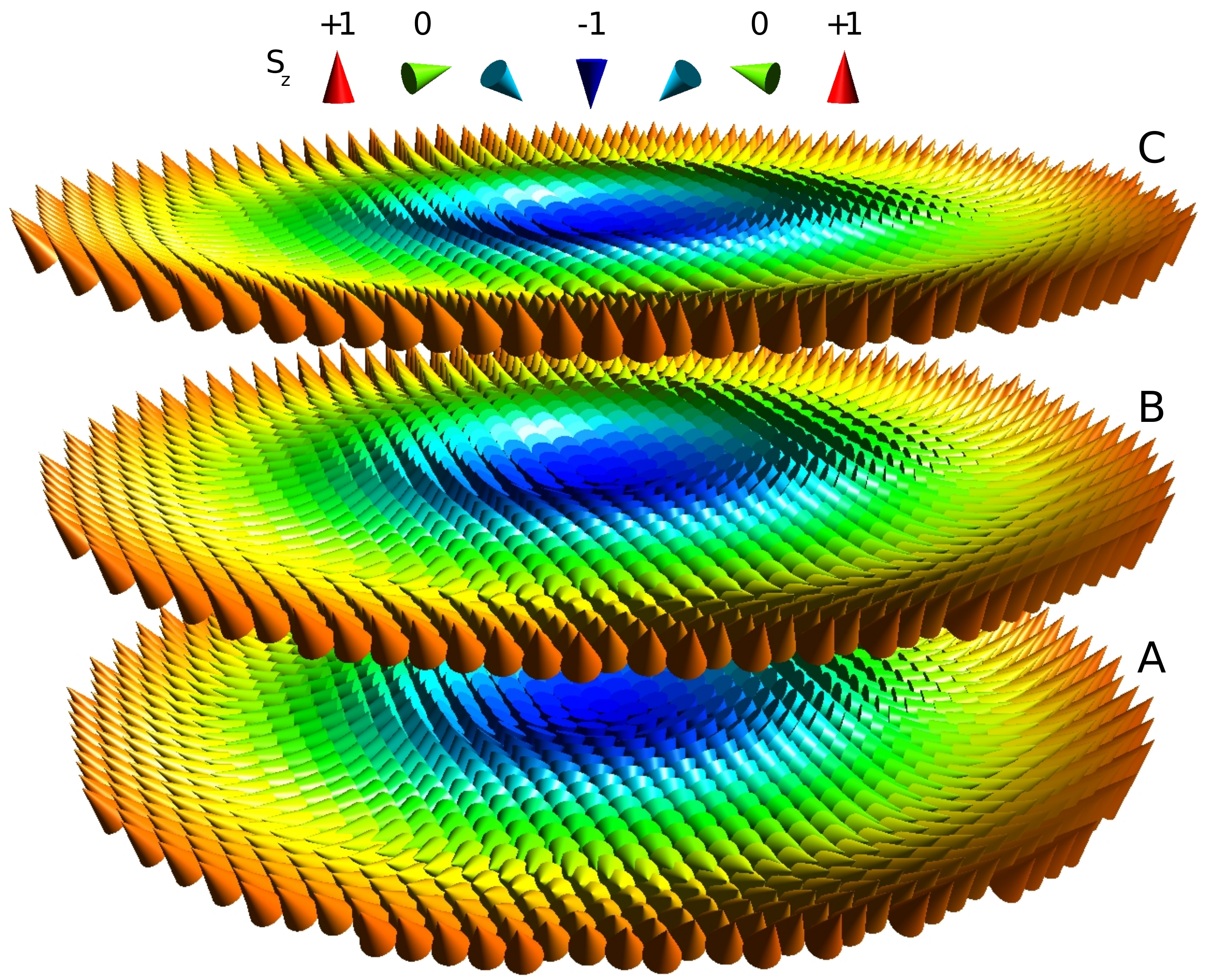}
\caption{Oblique view of the 3ML skyrmion structure, with color-coded spin directions. The monolayer-resolved DMI (shown in the left) corresponds to 3ML Co on Pt. The remaining parameters are taken to be $J=10$ meV per bond and $K=0.1$ meV per atom. An fcc stacking sequence is used, where interlayer distances are not drawn to scale.
\label{fig0}}
\end{center}
\end{figure}

In this work we consider a 0.6 nm thick cobalt (Co) film with perpendicular magnetic anisotropy on a platinum (Pt) substrate, without or with a capping layer of a different material (MgO, Ir, or other). The magnetic layer is composed of three atomic monolayers (MLs) with an fcc stacking sequence and a lattice constant of $a=2.51\AA$, as shown in Fig.~\ref{fig0}. We describe the magnetic material by a standard effective Hamiltonian
\begin{eqnarray}
H=&&-J\sum_{<i,j>}\bm{S}_i\cdot\bm{S}_j \nonumber\\ &&- \sum_{<i,j>}\bm{D}_{ij}\cdot (\bm{S}_i\times\bm{S}_j) - K\sum_i (S_i\cdot\bm{\hat{z}})^2 \nonumber\\ &&- \frac{\mu_0}{8\pi}\sum_{i,j\neq i} \frac{3(\bm{\mu}_i\cdot\bm{u}_{ij})(\bm{\mu}_j\cdot\bm{u}_{ij})-\bm{\mu}_i\cdot\bm{\mu}_j}{r_{ij}^3}
\label{heis}
\end{eqnarray}
within the Heisenberg model. $\bm{S}_i=\bm{\mu}_i/\mu$ is a three-dimensional unit vector representing the orientation of the effective magnetic moment $\bm{\mu}_i$ localized on atomic sites, where $\bm{\mu}_i=2.1\mu_B$ and $\mu_B$ is the Bohr magneton. $J$ is the effective nearest-neighbor exchange integral and $\bm{D}_{ij}$ is the DM vector arising from a three-site indirect exchange mechanism that couples two atomic spins to a neighbor atom with large spin-orbit coupling~\cite{Fert1980}. Such interface-induced DMI vectors are perpendicular to the unit vector $\bm{\hat{u}}_{ij}$ connecting $\bm{S}_i$ and $\bm{S}_j$ within the interface plane, namely $\bm{D}_{ij}=D(\bm{\hat{u}}_{ij}\times\bm{\hat{z}})$. $K$ describes the uniaxial perpendicular magnetic anisotropy with $\bm{\hat{z}}$ normal to plane. The last term in Eq. (\ref{heis}) corresponds to the dipole-dipole interaction between two atomic spins, where $r_{ij}$ is the distance between sites $i$ and $j$.

Since the dipolar coupling mainly acts as a shape anisotropy~\cite{Bogdanov}, it is possible to include its effect by reducing the anisotropy constant in the Hamiltonian~\cite{Rohart,Lobanov}. This has been carefully examined in the current work for few-atom thick films by directly calculating the dipolar interactions in the full neighborhood (i.e. all spin pairs). The massive computational effort involved, with over one billion atomic spin pairs, is tackled by graphics processing units (GPUs) for this part of the calculations. Similarity in the obtained results point to the same conclusions as in the monolayer case, so we consider a reduced anisotropy constant which effectively mimics the inclusion of dipolar interactions, muck like in Refs.~\onlinecite{Rohart,Lobanov}. References to anisotropy in the text are in terms of the effective anisotropy, where the dipolar term is implicitly included.

The considered sample is delimited by free boundary conditions and the size is chosen to roughly match the width of skyrmion-based racetrack memories proposed to date~\cite{Fert}. In this context, free boundary conditions means that the exchange and DM bonds are ignored between a boundary atom and anything that is outside of the finite geometry. This approach leads to similar magnetization profiles (canting of spins along the boundary) as those found in Refs.~\onlinecite{Rohart1,Sanchez}, which have included explicit corrections in the boundary conditions to account for DM interactions. A lattice of $130\times 130\times 3$ spins is used, which roughly corresponds to a $32\times 28\times 0.6$ nm nanostructure.  Simulations are performed at zero temperature and in absence of magnetic fields. The exchange stiffness ($J=29$ meV per bond), DM interaction ($D=1.5$ meV per bond), and effective anisotropy ($K=0.276$ meV per atom) are taken from experimental works on Co/Pt samples, as listed in~\cite{Rohart}. For particular cases considered in this work $D$ is retrieved from ab initio calculations~\cite{Yang,Boulle}, where DM interaction is inhomogeneous over the Co monolayers (see Fig.~\ref{fig0}), and can exhibit fairly nontrivial distribution depending on the chosen capping layer on top of Co. For the anisotropy and exchange interaction, however, we kept the parameter values homogeneous throughout the magnetic material. This is mainly caused by the lack of available information on layer-resolved interactions, but it also facilitates the understanding of our results (since we avoid competing interactions between the known monolayer-resolved DMI and the arbitrary variation of other parameters). Note that different $J$ and $K$ are necessary for stabilizing skyrmions with monolayer-dependent DMI, compared to the case of uniform (average) $D$ from Ref.~\cite{Rohart}. In our simulations, both interlayer and intralayer bonds are considered for the exchange and DM interactions using only the nearest neighbors. To stabilize the skyrmion, the system is numerically relaxed with a steepest descent method after initializing a circle of spins (with radius fixed to 4 nm) which point in the opposite direction of magnetization~\cite{Sampaio}. This hard circular domain ensures convergence to a (metastable) Sk state if such a solution exists, since the spin configuration must pass through it before reaching the FM state.

\section{Stability of a skyrmion\label{stab}}
\subsection{Three monolayers Co on Pt}
We start the analysis by considering the case of constant (uniform) DM interaction (DMI) in the 3ML Co film on Pt. The skyrmion found is similar to the one in a Co monolayer with same DMI~\cite{Rohart,Lobanov}, where its energy of $1580$ meV roughly matches the $500$ meV for a monolayer skyrmion considering there are three times more atoms in an 3ML fcc structure than in a two dimensional triangular lattice, albeit with a slightly higher diameter of $6.4$ nm instead of $4.6$ nm in the monolayer case. The diameter is calculated as the sum of inter-atomic distance between spins with $\bm{S}_i\cdot\bm{\hat{z}}\leq 0$. However, assuming a homogeneous effective DMI for few monolayers is highly unrealistic since the DMI strength $D$ should be inversely proportional to the thickness of the magnetic layer~\cite{Cho,Nembach}. A more accurate approach is to obtain the monolayer-resolved $D$ from first principles~\cite{Yang,Boulle} and incorporate it in our atomistic spin model. Here $D$ takes a large postive value in the first monolayer ($2.25$ meV per bond) and weakly proliferates away from the Co/Pt interface ($-0.3$ and $-0.4$ meV per bond for the middle and top layers, respectively). Although Ref.~\onlinecite{Yang} neglects interlayer contributions as being too small, we include them in our calculations as the average between DMIs in successive monolayers. In this approach we find that the skyrmion is no longer stable for the conditions defined earlier~\cite{Rohart}. Instead, a full sweep of the relevant parameters is needed to find a parametric domain where skyrmions are stable.

\begin{figure}
\begin{center}
\includegraphics[width=\columnwidth]{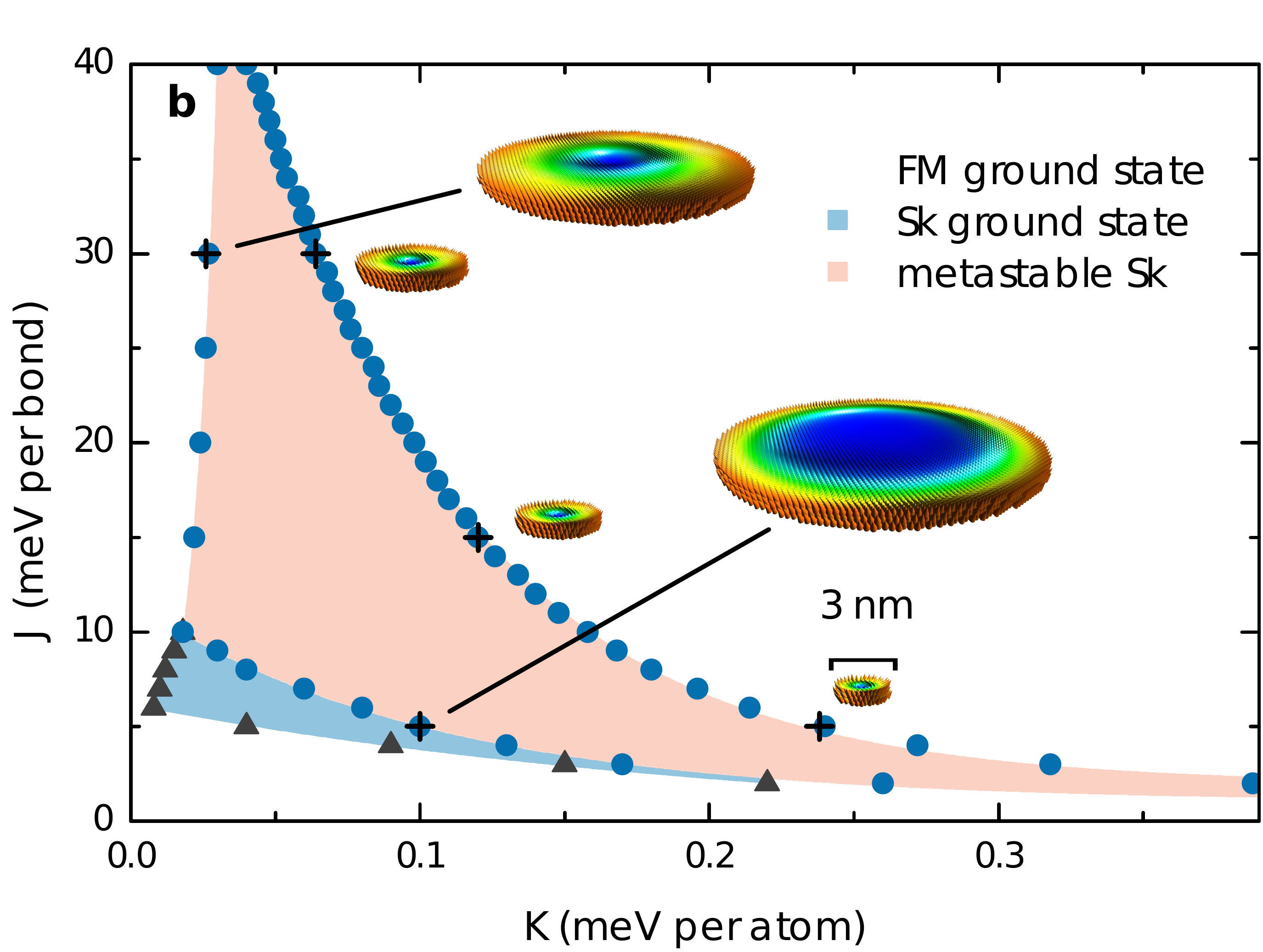}
\caption{%(a) Layer resolved DMI strength for the 3ML Co on Pt, retrieved from ab initio calculations~\cite{Yang,Boulle}. (b)
Magnetic state diagram at zero temperature and in absence of magnetic field, as a function of anisotropy and exchange interaction, for DMI taken as in Fig.~\ref{fig0}. Data points (circles and triangles) are retrieved from numerical simulations. The transition boundaries separate regions of metastable skyrmions (shaded in red) from the uniform (FM) ground state (white), and the skyrmion ground state (shaded in blue). Skyrmions are shown for different locations in the parameter space, denoted by crosses. 
\label{fig1}}
\end{center}
\end{figure}

For the case of monolayer-resolved DMI in a 3ML Co film, we constructed a magnetic state diagram as a function of magnetic anisotropy, $K$, and exchange coupling, $J$. Fig.~\ref{fig1} shows that metastable skyrmions can form in a significant portion of the parameter space, albeit at zero temperature. The rest of the diagram corresponds to the uniform (FM) ground state. We find that skyrmions can only exist for anisotropies that are much lower than expected. In particular, the Sk state survives when $K$ is almost an order of magnitude smaller than in the usual case of homogeneous DMI ($K=0.276$ meV per atom). A plausible explanation is that vertical dilution of DMI in Co monolayers reduces the ability for a skyrmion to form through the entire depth of the magnetic material. Atomic layers further from the Pt substrate will have weak DMI that cannot sustain a stable skyrmion structure, which then propagates into the lower layers through the exchange coupling. Strong ferromagnetic ordering therefore takes over, whereas a low exchange coupling would restore the stability of skyrmions. We also note the inverse relation between the effects of anisotropy and the exchange interaction on skyrmion stability, well characterized by the transition lines and the shape of the stability domain of the skyrmion in Fig.~\ref{fig1}.

\begin{figure}
\begin{center}
\includegraphics[width=\columnwidth]{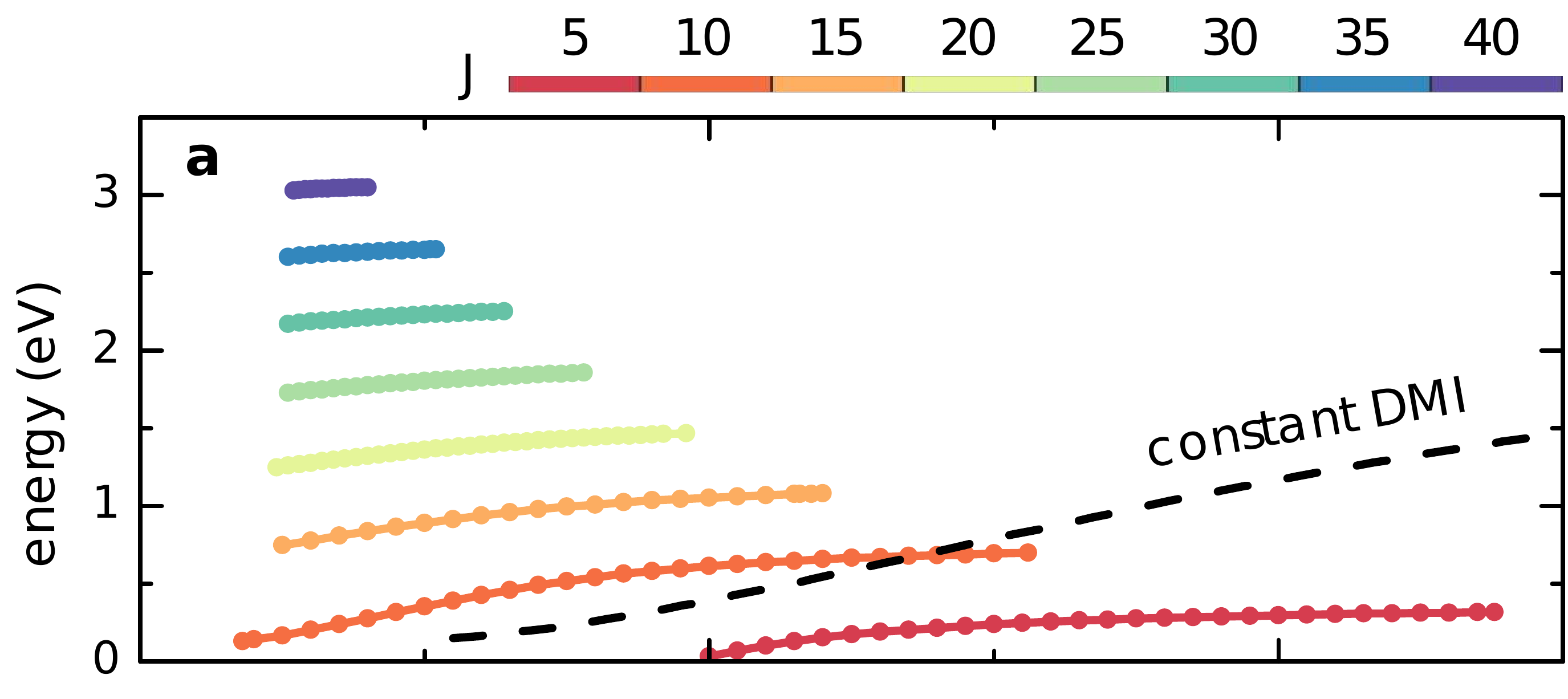}
\includegraphics[width=\columnwidth]{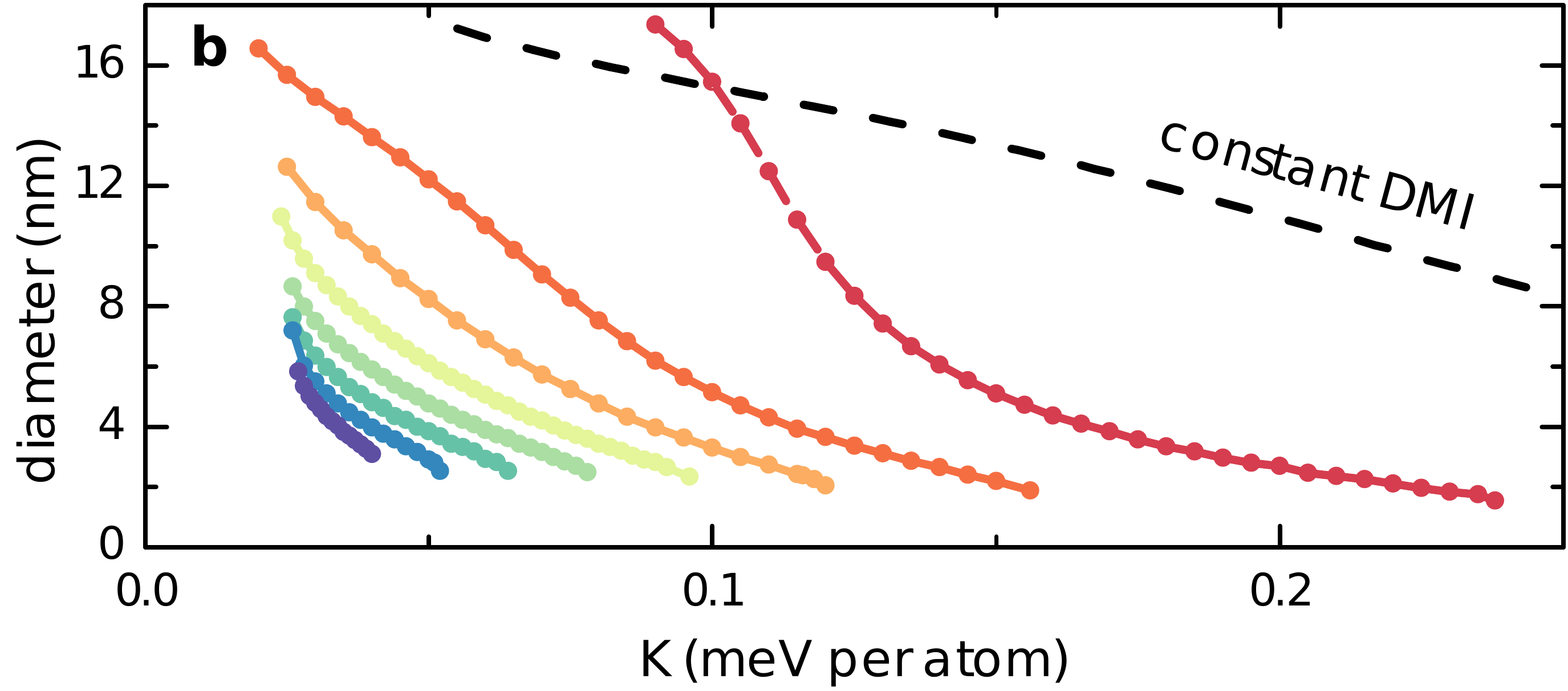}
\caption{(a) Skyrmion energy and (b) skyrmion size as a function of magnetic anisotropy for fixed values of the exchange coupling, $J$. Dotted line corresponds to the case of homogeneous DMI, with parameter values taken from Ref. \onlinecite{Rohart}. \label{fig2}}
\end{center}
\end{figure}

Fig.~\ref{fig2}(a) shows that the Sk energy (calculated as the energy difference between the Sk and FM state) is higher for stronger exchange interactions and increases with increasing magnetic anisotropy. For strong exchange ($J=30$ meV per bond) the skyrmion has an average energy of $2.2$ eV compared to $0.5$ eV for weak exchange ($J=10$ meV per bond). It can also be seen that the variation of energy as a function of $K$ difference is only $80$ meV in the former case but as high as $202$ meV in the latter. This demonstrates that the anisotropy has a more pronounced effect on the skyrmion energy for weak exchange interactions. Energetically favorable configurations, that are closer to the ground state, can thus be obtained by decreasing either of the two parameters. The size (diameter) of the skyrmion core is inversely proportional to the anisotropy, which lies in correspondence with its energy [Fig.~\ref{fig2}(b)]. As anisotropy decreases the skyrmion grows in size until the remaining outer spins are flipped to the uniform ground state (all spins down), or until the skyrmion slides outside from the boundary. For large $K$ and lower $J$ the skyrmion shrinks in size and its energy increases until it overcomes the energy barrier and collapses to the uniform ground state with all spins up. It is worth noting that the skyrmions found here can be as small as $1-2$ nm in diameter or as large as $\sim18$ nm for low exchange constants. For high $J$, the skyrmions never surpass $10$ nm in diameter [see exemplified cases in Fig.~\ref{fig1}]. This implies that the constraining geometry of finite systems can have a strong effect on skyrmion size and energy, especially in the presence of strong exchange coupling~\cite{Mulkers}.

\begin{figure}
\begin{center}
\includegraphics[width=\columnwidth]{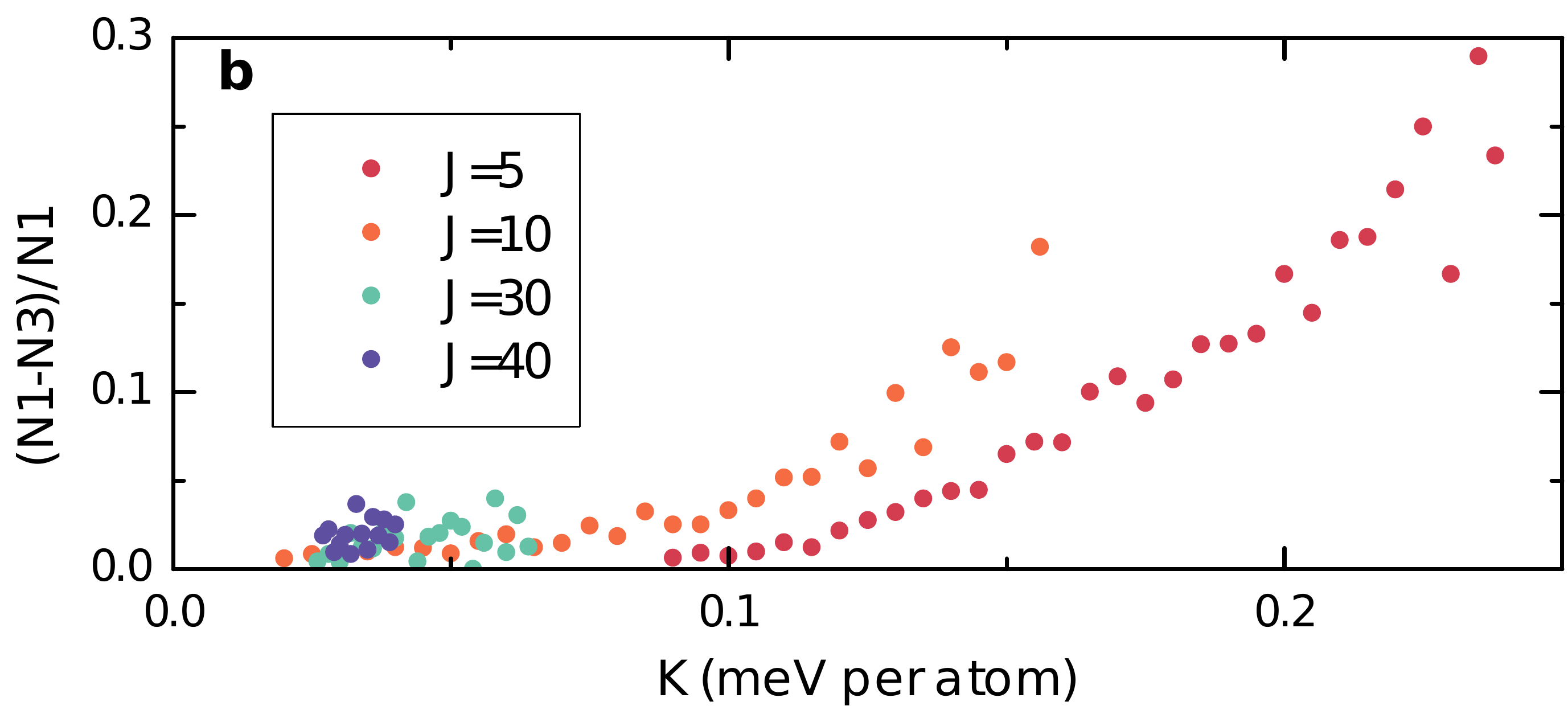}
\caption{%(a) Oblique view of the 3ML skyrmion structure with red arrows pointing up and blue arrows pointing down. The parameters are taken to be $J=10$ meV per bond, $K=0.1$ meV per atom, and $D$ is retrieved from ab initio calculations [Fig.~\ref{fig1}(a)]. Interlayer distances are not drawn to scale.  The skyrmion diameter in each monolayer is $4.7$ nm, $4.5$ nm, and $4.5$ nm from bottom (closest to substrate) to top, respectively. (b)
Relative difference between the number of core spins in the first and the third monolayer [$(N_c^{(1)}-N_c^{(3)})\big/N_c^{(1)}$)] as a function of magnetic anisotropy for weak and strong exchange interactions.\label{fig3}}
\end{center}
\end{figure}

It is informative to compare the skyrmion structure in our approach to that in the case of homogeneous DMI. The skyrmion energy in the latter case is $1.59$ eV, and skyrmion core has a diameter of $6.4$ nm in all three monolayers. In Fig.~\ref{fig0} we have shown that for a 3ML sample with monolayer-resolved DMI, the Sk structure persists through the entire magnetic material. The skyrmion core, however, can shrink for monolayers far from the Co/Pt interface, since the DMI strength decreases vertically in the magnetic material. In Fig.~\ref{fig3} we observe for what range of parameters are there significant variations in skyrmion size between individual monolayers. The figure shows that the skyrmion diameter depends heavily on the strength of exchange interactions. While fluctuations in skyrmion size between monolayers are negligible for strong exchange (less than $5\%$ regardless of the anisotropy, which amounts to only a few spins), for weak $J$ they vary from small to large as we increase the anisotropy. A plausible explanation for this behavior is that strong exchange interactions prevent distortion of the skyrmion structure across different monolayers. We can therefore expect that interlayer variations are not important in the few ML Co on Pt system considered here.

\begin{figure}
\begin{center}
\includegraphics[width=\columnwidth]{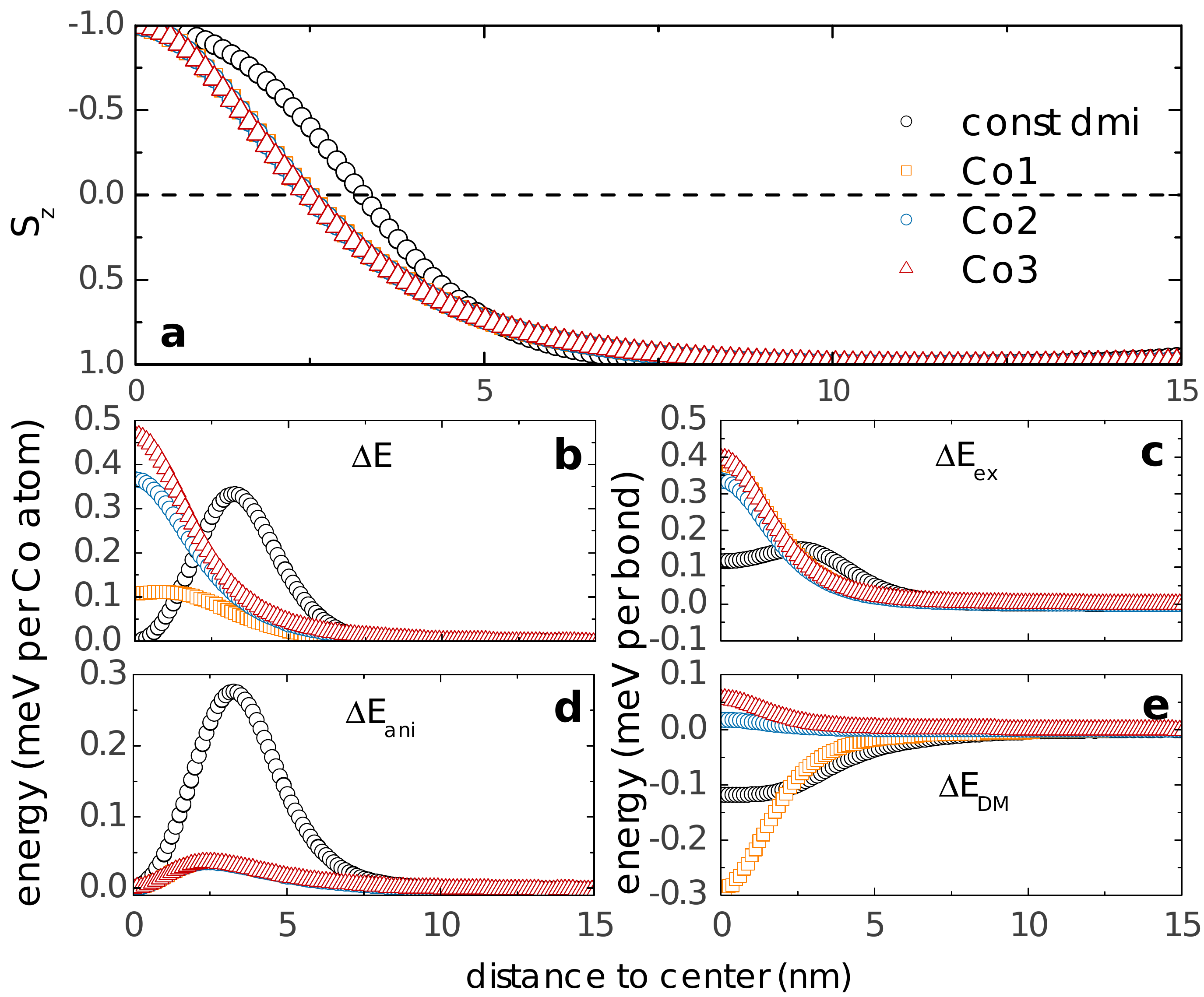}
\caption{(a) Line profile of an individual skyrmion for homogeneous DMI and for monolayer-resolved DMI [see Fig.~\ref{fig0}]. Strong exchange coupling is assumed ($J=30$ meV per bond) and the anisotropy is chosen to be $K=0.036$ meV per atom. The labels refer to the individual Co monolayers that are closest (Co1), in between (Co2), and furthest (Co3) from the Pt substrate. (b) Total energy difference (exchange, DM, and anisotropy) per Co atom with respect to the uniform FM state as a function of distance from the skyrmion center. (c-e) Energy difference due to exchange interaction, magnetic anisotropy and DM interaction, respectively. \label{fig4}}
\end{center}
\end{figure}

Fig.~\ref{fig4}(a) reveals a change in skyrmion profile compared with that for homogeneous DMI. In particular, the skyrmion is more extended in the sense that spins rotate to the uniform FM state at a slower spatial rate while the core spins ($S_z\leq 0$) have smaller magnitudes. However, this could be a result of the different anisotropies used in the two cases which induce different skyrmion sizes. The fact that the profile is identical throughout all three Co monolayers matches our previous results regarding skyrmion diameter throughout the magnetic structure for strong exchange couplings [see Fig.~\ref{fig3}]. The total energy contribution per Co atom appears to be significantly different. For the homogeneous DMI case the energy profile (energy difference per Co atom or bond as a function of distance from skyrmion center) has a maximum ($E>0$) followed by a minimum ($E<0$) as expected from literature~\cite{Dupe,Dupe1}. In our approach there are three distinct energy profiles for different monolayers, which correspond to the different DMIs in the three layers. The monolayer closest to the surface has an energy profile similar to one found in the homogeneous case, but the energy profiles of the two monolayers further above follow a monotonic decrease.

These energy profiles can be understood by analyzing individual contributions of different magnetic interactions, plotted in Fig.~\ref{fig4}(c-e). The two curves in Fig.~\ref{fig4}(c) distinguish the stronger local variation of the exchange energy in the central monolayer compared to the top and bottom ones. There is a peak in exchange energy for the homogeneous case as the spins rotate from the skyrmion core to ferromagnetic ordering. This is not observed for the monolayer-resolved DMI because the core is already too small to have spins aligned in the center. In that case the DM energy takes three different profiles which correspond to the dilution of DMI in three Co monolayers [according to Fig.~\ref{fig0}]. The difference in total energy profiles arises from the cancellation of contributions of exchange and DM interactions in the first monolayer, but not in the other two due to low DM energy there. In the homogeneous case the anisotropy energy has a significantly stronger effect on the total energy per Co atom compared to the monolayer-resolved DMI [see Fig.~\ref{fig4}(d)]. This can be explained by the higher $K$ needed to stabilize a skyrmion with homogeneous DMI [see Fig.~\ref{fig2}].

\subsection{Three monolayers Co in a spin-orbit sandwich}
\begin{figure}
\begin{center}
\includegraphics[width=\columnwidth]{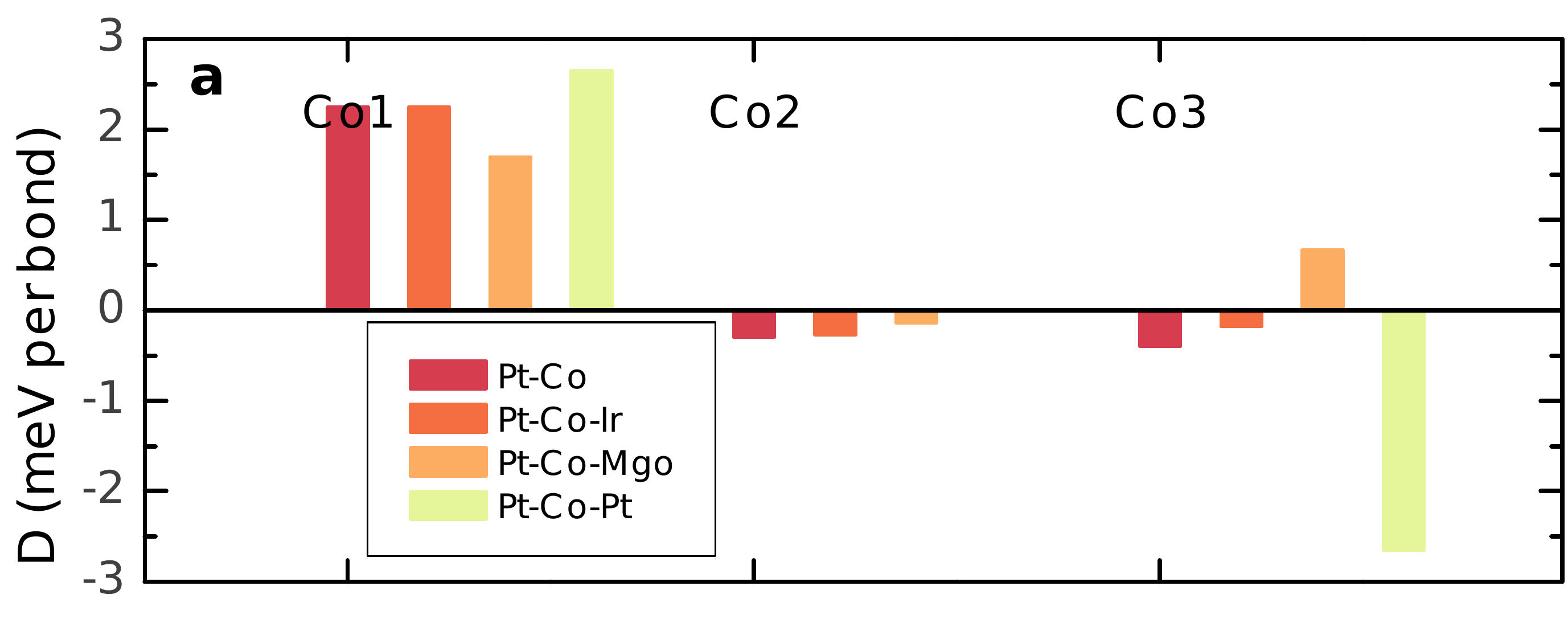}
\includegraphics[width=0.49\columnwidth]{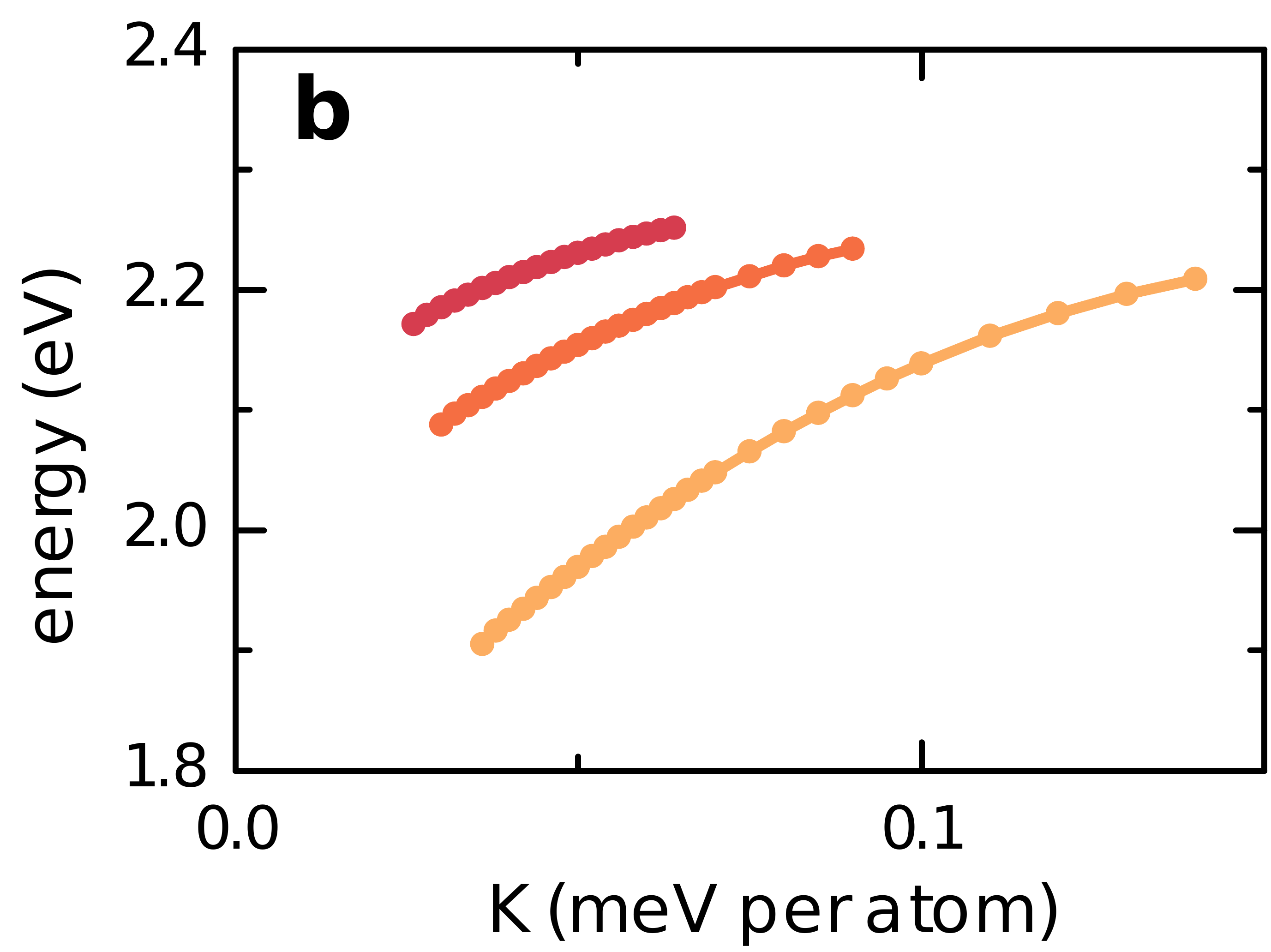}
\includegraphics[width=0.49\columnwidth]{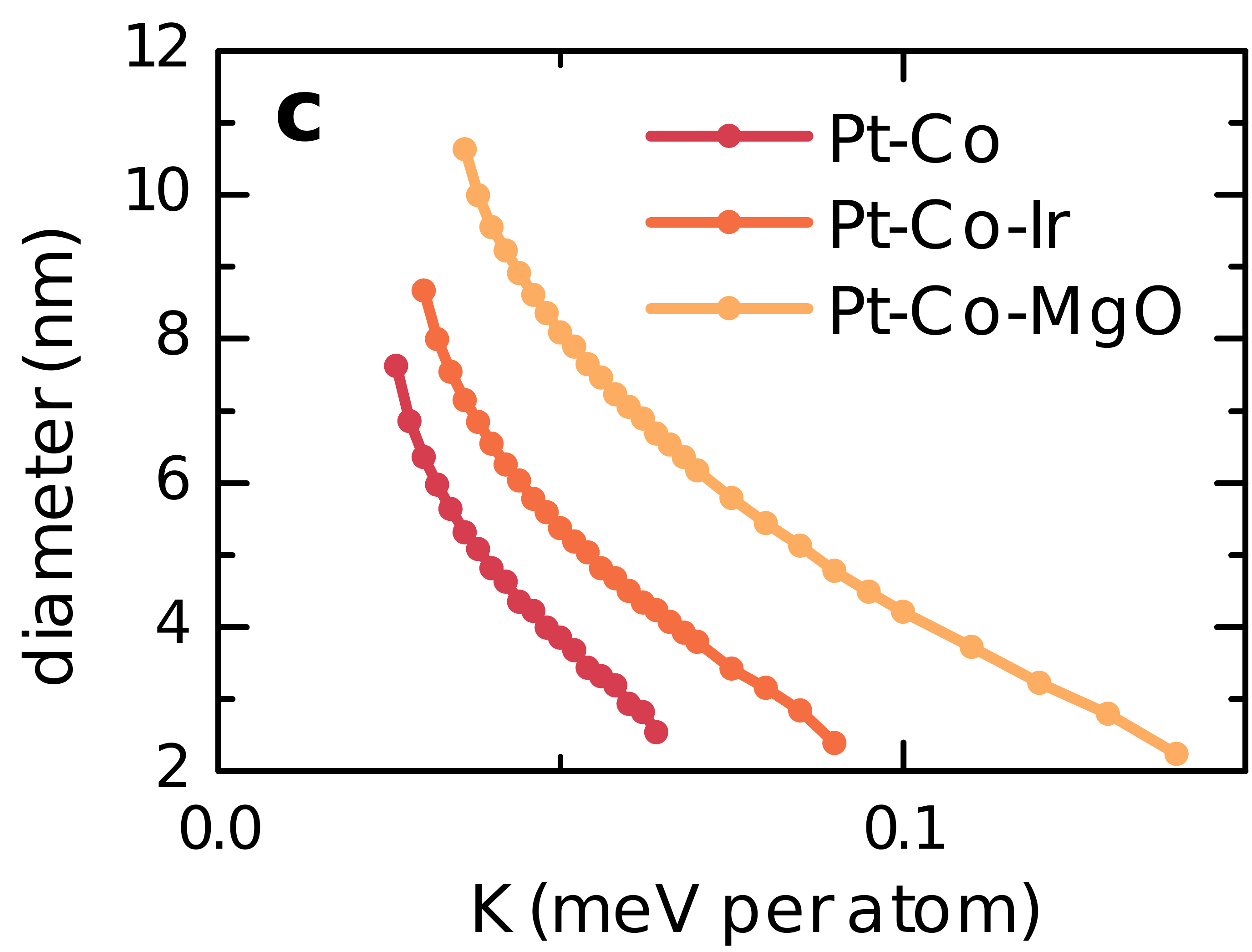}
\caption{(a) Layer resolved DMI strength for the 3ML Co in considered sandwich multilayers, compared to the case of epitaxial 3ML Co on Pt, retrieved from ab initio calculations~\cite{Yang,Boulle}. (b) Skyrmion energy and (c) skyrmion size (diameter) as a function of magnetic anisotropy for strong exchange coupling ($J=30$ meV per bond).
\label{fig5}}
\end{center}
\end{figure}

In what follows, we consider samples where the magnetic film is sandwiched between materials that provide (a)symmetric spin-orbit coupling (Pt-Co-MgO~\cite{Boulle}, Pt-Co-Ir~\cite{Moreau,Yang}, and Pt-Co-Pt~\cite{Yang}). The choice of materials is guided by recent experiments where skyrmions were stabilized at room-temperature~\cite{Boulle,Moreau} and by ab initio predictions for DMI in ferromagnetic/heavy metal bilayers~\cite{Yang}. This class of samples allows for relatively large DMI to be achieved, which is clearly favorable for skyrmions. In Fig.~\ref{fig5}(a), we show the comparative analysis of monolayer-resolved DMI in sandwiched 3ML Co samples and the previously considered 3ML Co on Pt. DMI remains largest in the first Co monolayer, mainly stemming from the Pt-Co interface~\cite{Boulle}. In a Pt-Co-Ir sandwich, the DMI for Co-Ir interface is weaker than for the third Co monolayer for 3ML Co on Pt, which reduces the negative value of DMI in the central Co monolayer and increases the total DMI. In a Pt-Co-MgO case, DMI at the Co-MgO interface has the same sign as that of Pt-Co, which enhances the total DMI and further reduces the negative DMI in the central monolayer. Pt-Co-Pt samples would ideally have symmetric interfaces, which leads to no DMI in the central monolayer and (in practice nearly) zero overall DMI~\cite{Yang}.

With this input for monolayer-resolved DMI, we present the properties of isolated skyrmions in the discussed hybrid multilayer structures, retrieved from atomic-spin simulations. Fig.~\ref{fig5}(b) shows that Sk energies have similar (monotonically) increasing profiles as a function of anisotropy as found in the 3ML Co on Pt case. For the assumed identical exchange coupling, skyrmions in Pt-Co-MgO samples have lower energies than those found in Pt-Co-Ir, which are in turn lower than in Pt-Co case. No skyrmion states were found in Pt-Co-Pt, for either strong or weak exchange interactions, because of the specific self-canceling DMI in that case. We also note that in multilayers the skyrmion stability is extended to higher anisotropy values, as expected from the increase in total DMI seen in Fig.~\ref{fig5}(a). In particular, the energy profile of Pt-Co-MgO extends significantly more to higher anisotropy since the DMI strength is positive in the last Co ML, contrary to negative values found in Co-Pt-Ir and Pt-Co cases. The range of possible skyrmion sizes also increases in multilayer structures [see Fig.~\ref{fig5}(c)]. For strong exchange coupling, the largest found skyrmions are $7$ nm in diameter for Pt-Co, $8$ nm for Pt-Co-Ir, and $10$ nm for Pt-Co-MgO, whereas the smallest ones are $2.3$ nm, $2$ nm and $1.8$ nm, respectively.

\begin{figure}
\begin{center}
\includegraphics[width=\columnwidth]{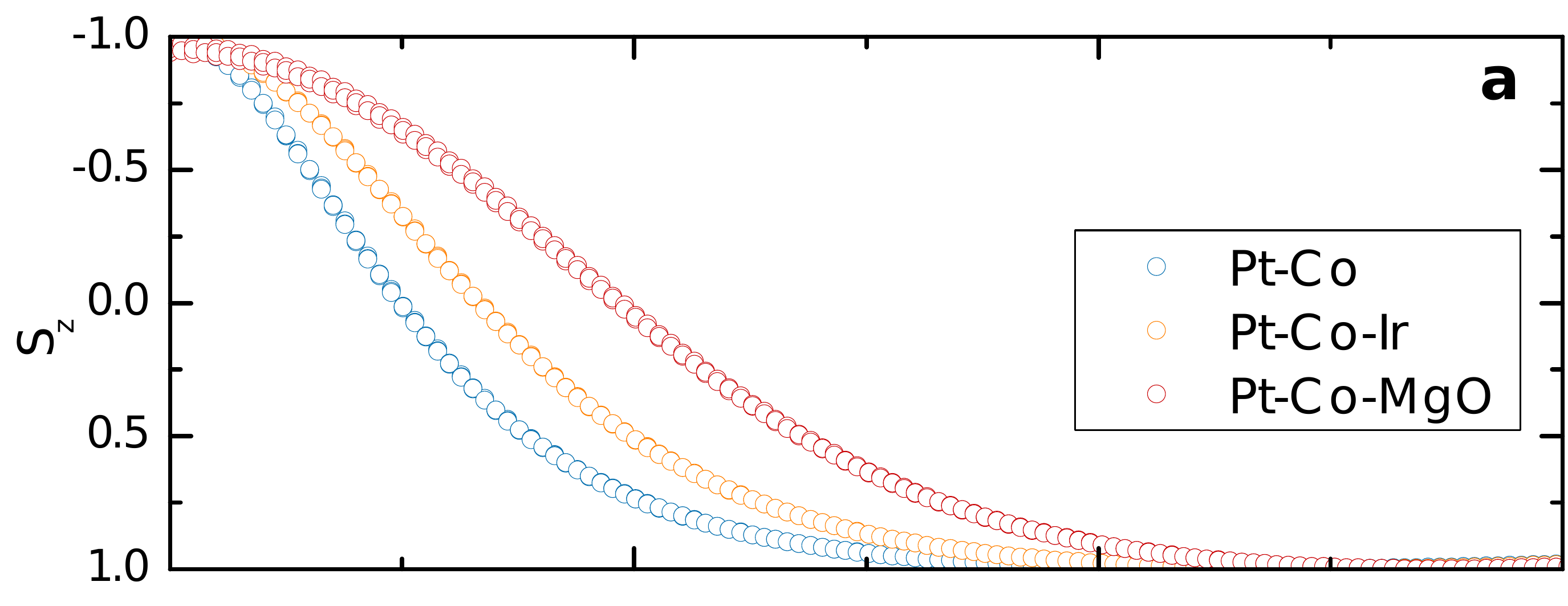}
\includegraphics[width=\columnwidth]{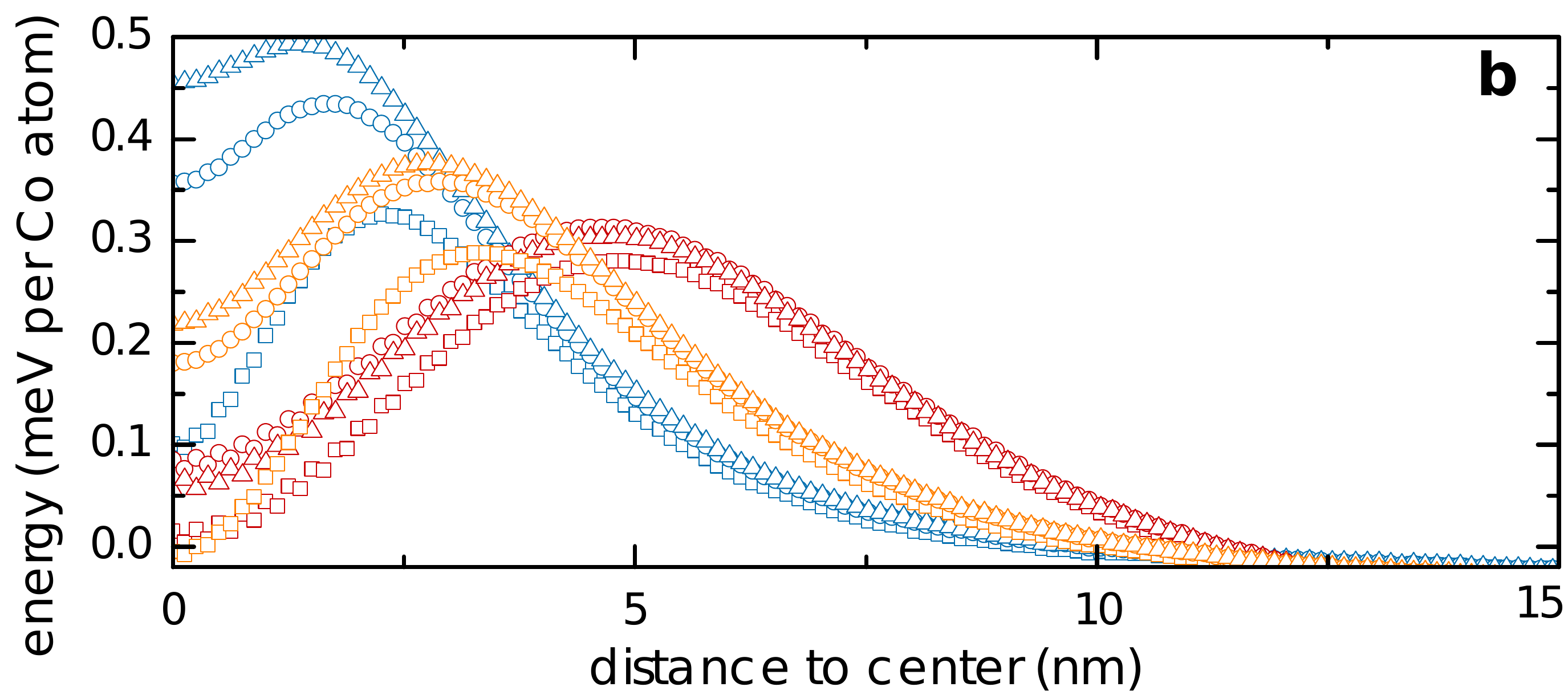}
\caption{(a) Line profile of an individual skyrmion for the 3ML Co in considered sandwich multilayers and the case of epitaxial 3ML Co on Pt. (b) The energy difference per Co atom with respect to the FM ground state as a function of distance from the skyrmion core. Strong exchange coupling is assumed ($J=30$ meV per bond) and the anisotropy is taken as $K=0.036$ meV per atom. The symbols correspond to the individual Co monolayers that are closest (squares), in between (circles), and farthest (triangles) from the Pt substrate.
\label{fig6}}
\end{center}
\end{figure}

In Fig.~\ref{fig6}, the total energy per Co atom is shown for the individual monolayers. In the Pt-Co-Ir case the energy profiles closely resemble those discussed for Pt-Co. In Pt-Co-MgO, however, the energy profiles differ significantly from latter two cases, as peaks in the energy profiles are shifted outward to about $5$ nm from the skyrmion core. The energy profiles of the three monolayers are much closer in magnitude than those found in Pt-Co and Pt-Co-Ir, mostly because the DMI has the same sign in the first and third monolayers  which results in more stable skyrmion structures. This suggests that the spatial profile of the skyrmion core is significantly different compared to other cases.

\begin{figure*}
\begin{center}
\includegraphics[trim = 0 0 0 0, clip, width=\textwidth]{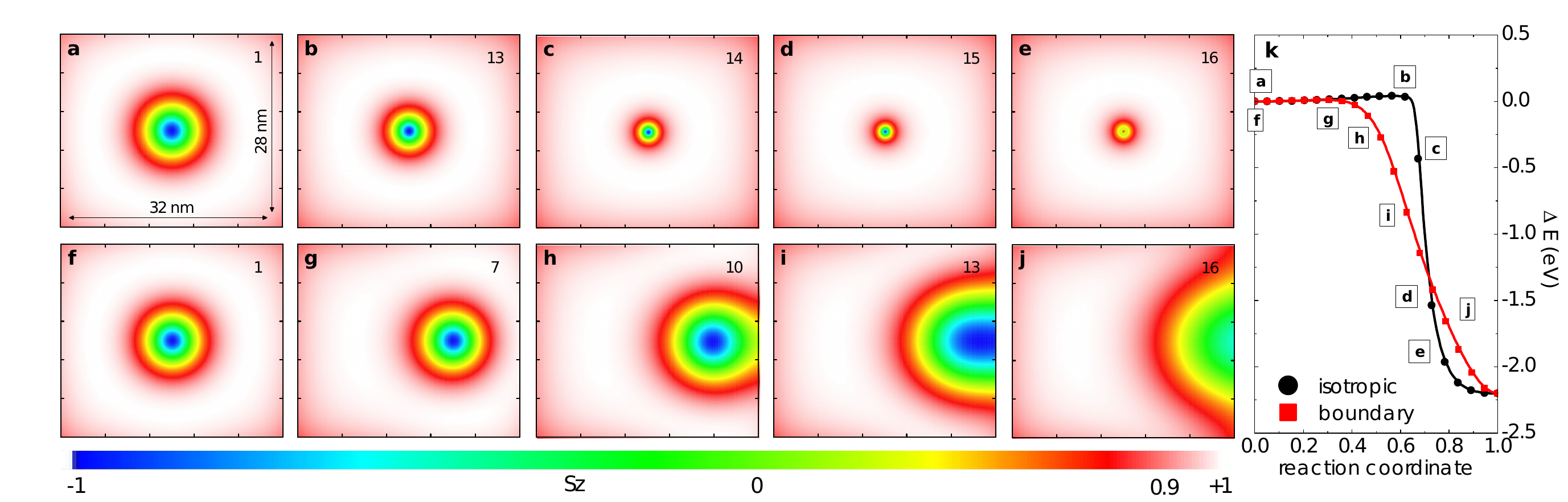}
\caption{Top view of magnetization ($S_z$) distribution for (a-e) isotropic and (f-j) boundary collapse paths of a skyrmion in 3ML Co on Pt, with monolayer resolved DMI. The skyrmion is initially $4.7$ nm in diameter, for taken $J=30$ meV per bond and $K=0.036$ meV per atom. $S_z$ is shown color coded from white to red to green to blue (shown above). The initial images (a, f) are followed by saddle points (b, g), and remaining images show configurations on the subsequent energy decline towards the FM ground state (not shown). Numbers on top right correspond to ordering of images in minimum-energy path curves for the isotropic (k) and boundary (l) collapse. The reaction coordinate is defined as the normalized (geodesic) displacement along the minimum energy path.
\label{fig7}}
\end{center}
\end{figure*}

\section{Energy paths for skyrmion collapse\label{coll}}
The thermal stability of magnetic skyrmions can be assessed by studying the mechanisms for skyrmion collapse and the associated energy barriers. Minimum energy path calculations are often used to find transition paths in atomistic systems, and were recently extended to magnetic systems as well~\cite{Dittrich,Bessarab,Rohart,Lobanov}. For example, Refs.~\onlinecite{Rohart,Lobanov} have used minimum energy paths to describe the collapse of isolated skyrmions. However, these works are restricted to a single atomic layer of magnetic material, and considered solely the isotropic (on itself) collapse of a skyrmion. Here we extend the material system to a more realistic few-monolayer case, sandwiched or not, laterally finite in size with either physical end of the ferromagnetic film or an interface where DMI vanishes (due to e.g. laterally finite heavy-metal layer). We consider two evolution paths the system can take in the process of skyrmion nucleation or annihilation: an isotropic transition where a symmetric rotation of the spins in the radial direction takes place~\cite{Rohart,Lobanov} and collapse at the boundary~\cite{CortesOrtuno}. We make use of the geodesic nudged elastic band method~\cite{Bessarab} to calculate minimum energy paths for an isolated skyrmion. After reproducing the test problems from Refs.~\onlinecite{Dittrich} and~\onlinecite{Bessarab}, we simulated the monolayer skyrmion of Ref.~\onlinecite{Rohart} and recovered the same energy barrier for isotropic collapse as in Ref.~\onlinecite{Lobanov}.

The minimum energy paths were calculated as follows. A set of intermediate replicas of the system (referred to as images) were generated to produce a discrete representation of the transition path~\cite{Lobanov}. Here, a total of $20$ images were considered with the initial and final points fixed to the Sk and FM state, respectively. The other $18$ images are energy minimized in a collective way using an iterative optimization procedure~\cite{Sheppard,Bessarab}. Relaxation occurs in the plane orthogonal to the tangent between successive images, which are connected by spring forces to ensure equal spacing along the transition path. The curved manifold of magnetic systems is taken into account by using a geodesic measure of distances and projecting path tangents and magnetic forces on its tangent space~\cite{Bessarab}. The highest image moves up the energy surface along the transition path using the climbing image algorithm~\cite{Henkelman}, and converges rigorously to a first-order saddle point. This can be verified by ensuring the energy gradient is sufficiently close to zero. Given the complexity of the energy surface, different stable paths can be obtained depending on the initial guess.

\begin{figure}
\begin{center}
\includegraphics[width=0.95\columnwidth]{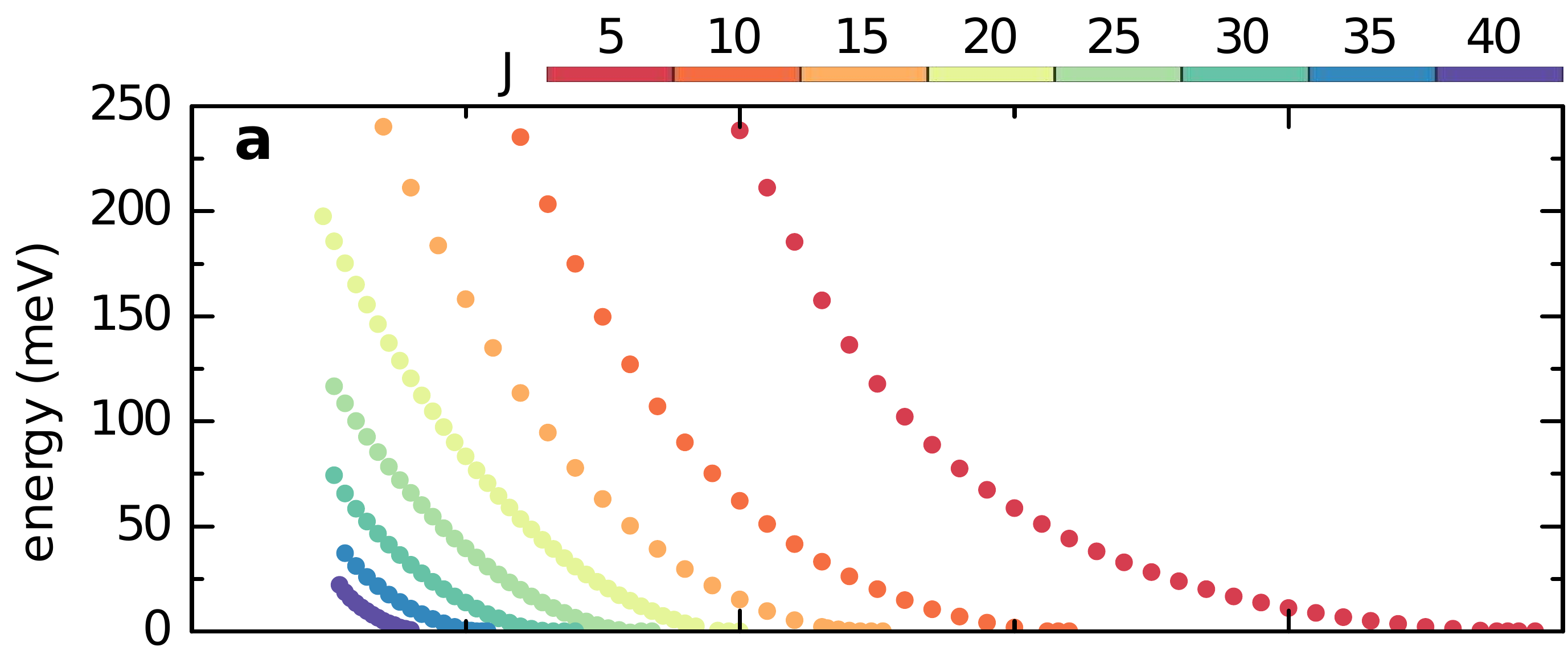}
\includegraphics[width=0.95\columnwidth]{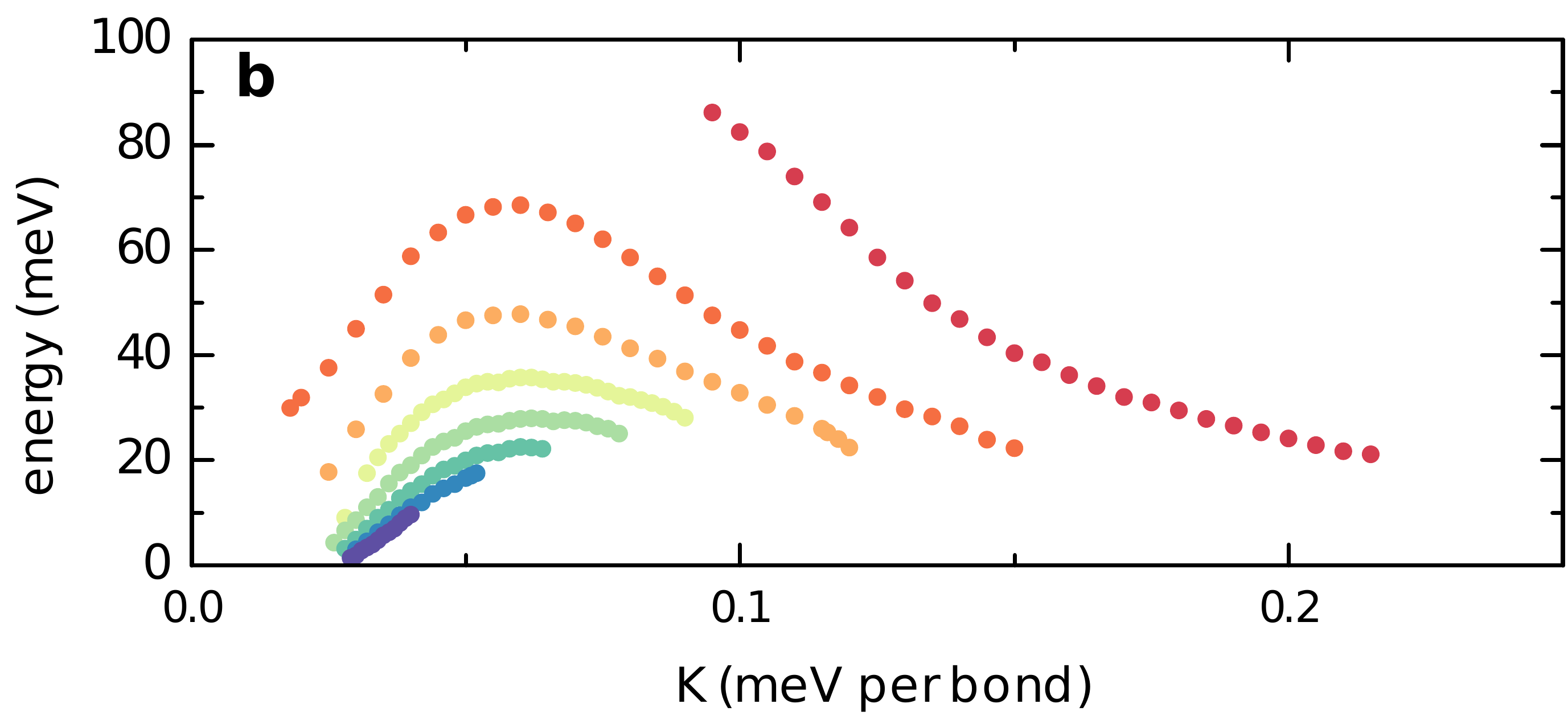}
\caption{Energy barriers for (a) isotropic and (b) boundary collapse of a skyrmion as a function of magnetic anisotropy. Each point corresponds to a unique energy barrier retrieved from the maximum (saddle point) in a minimum-energy path.\label{fig8}}
\end{center}
\end{figure}

\begin{figure}
\begin{center}
\includegraphics[width=\columnwidth]{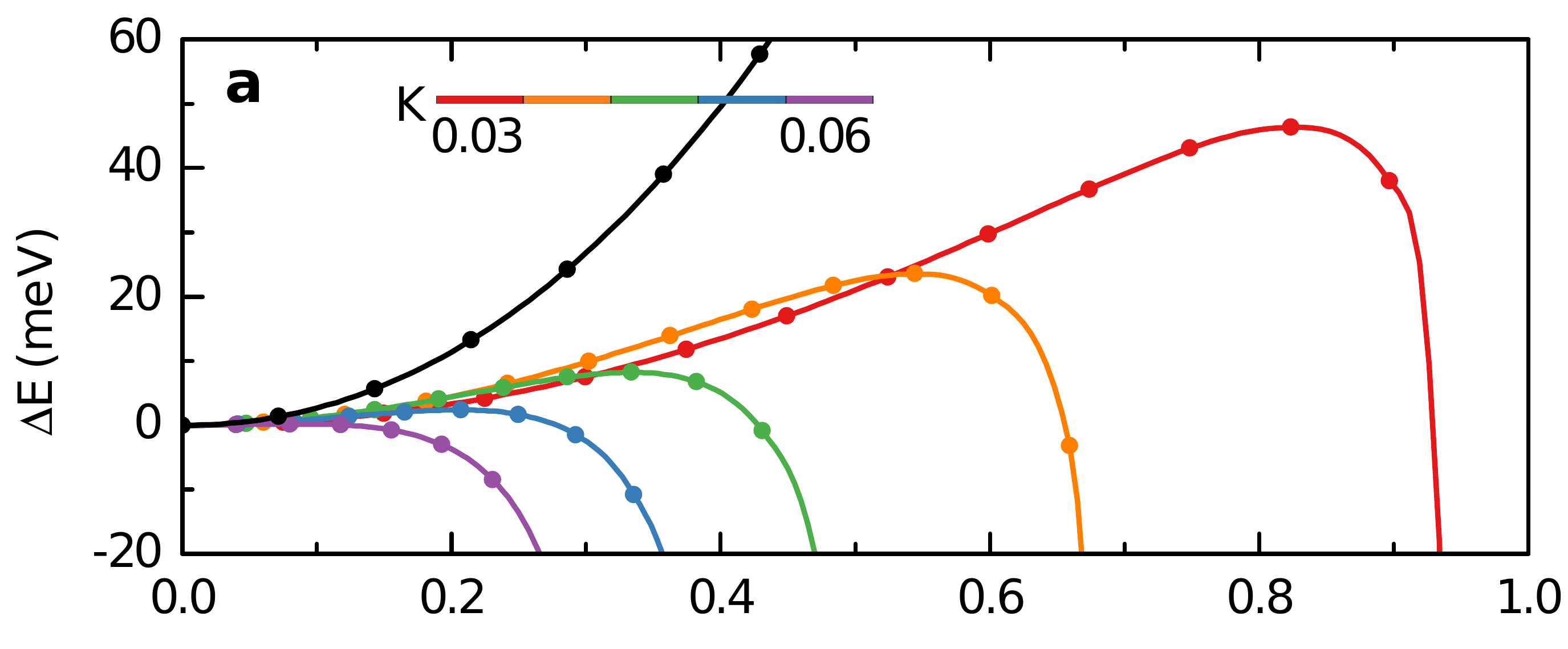}
\includegraphics[width=\columnwidth]{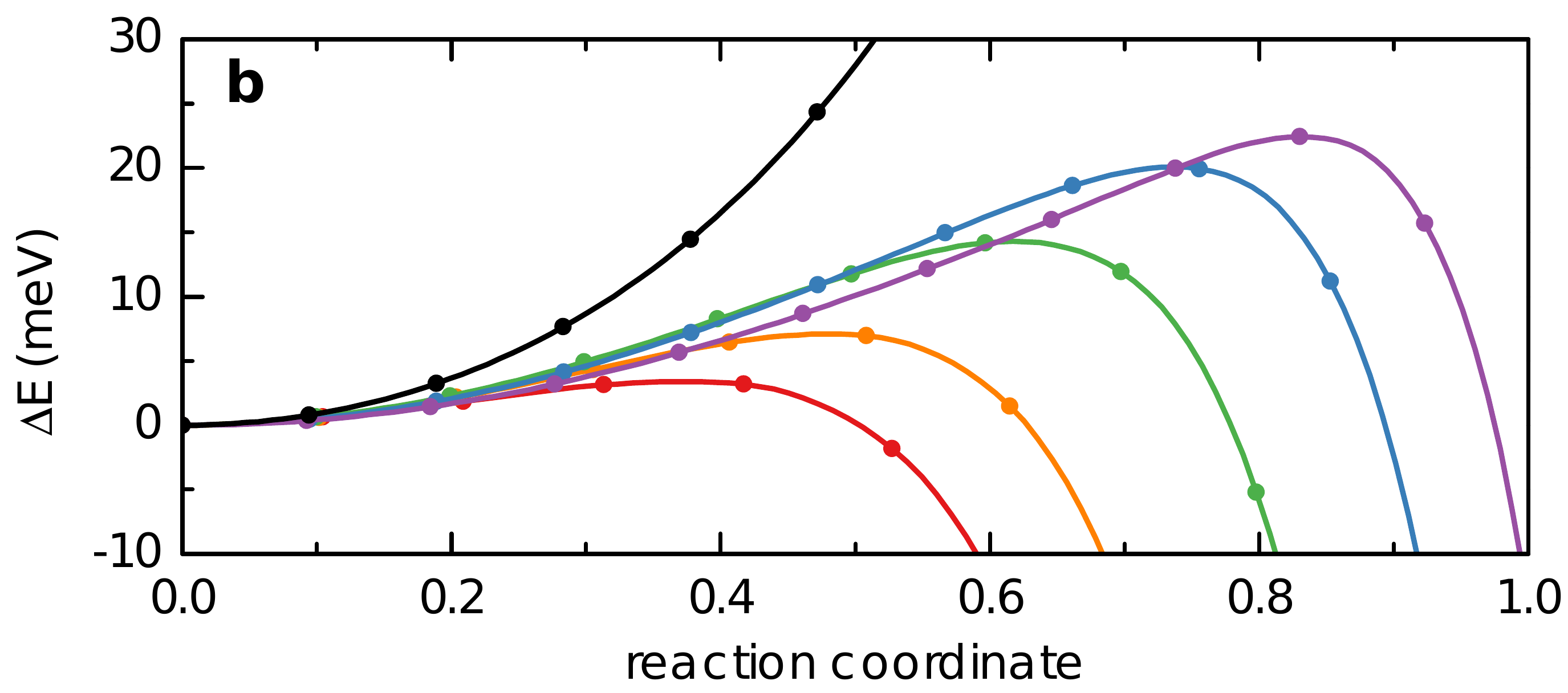}
\caption{Minimum-energy paths for (a) isotropic and (b) boundary skyrmion collapse in case of strong exchange coupling ($J=30$ meV per bond). Paths are shown for different magnetic anisotropies with respect to the FM ground state. Solid lines are cubic polynomial interpolations~\cite{Bessarab} of the discrete images (dots). Black lines correspond to the case of homogeneous DMI.\label{fig9}}
\end{center}
\end{figure}

\begin{figure}
\begin{center}
\includegraphics[width=\columnwidth]{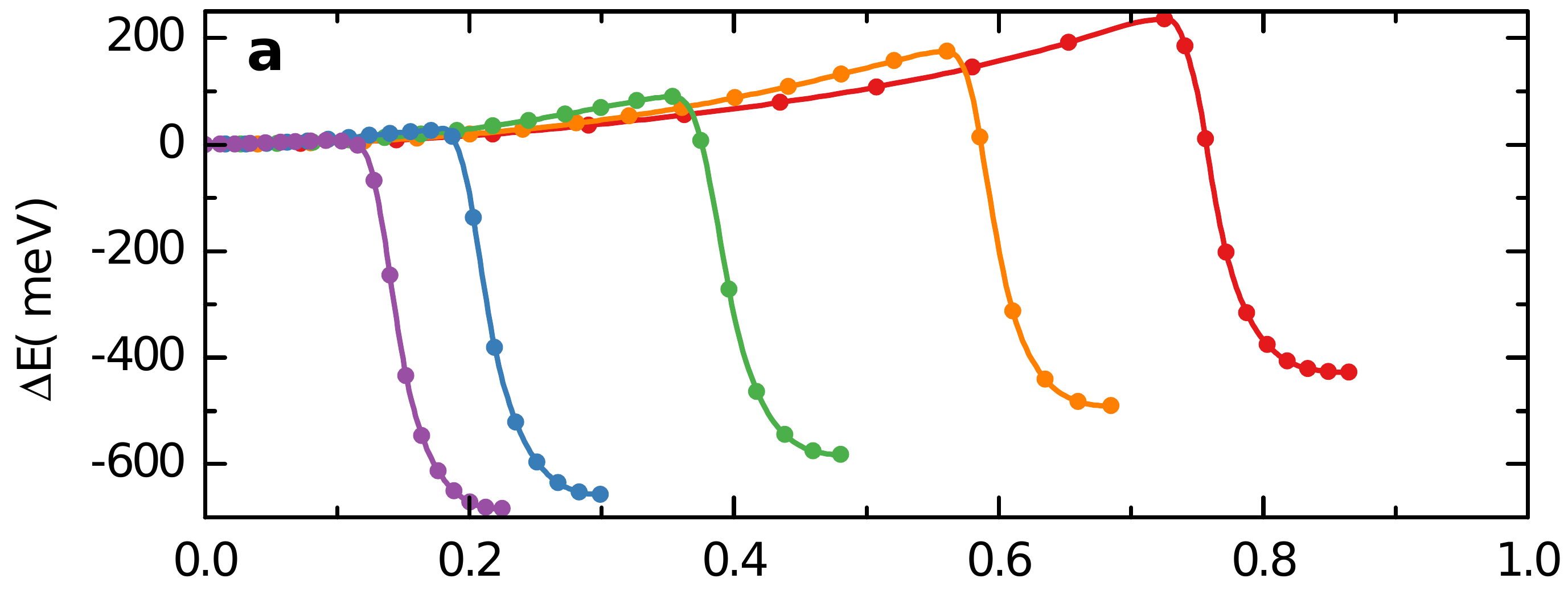}
\includegraphics[width=\columnwidth]{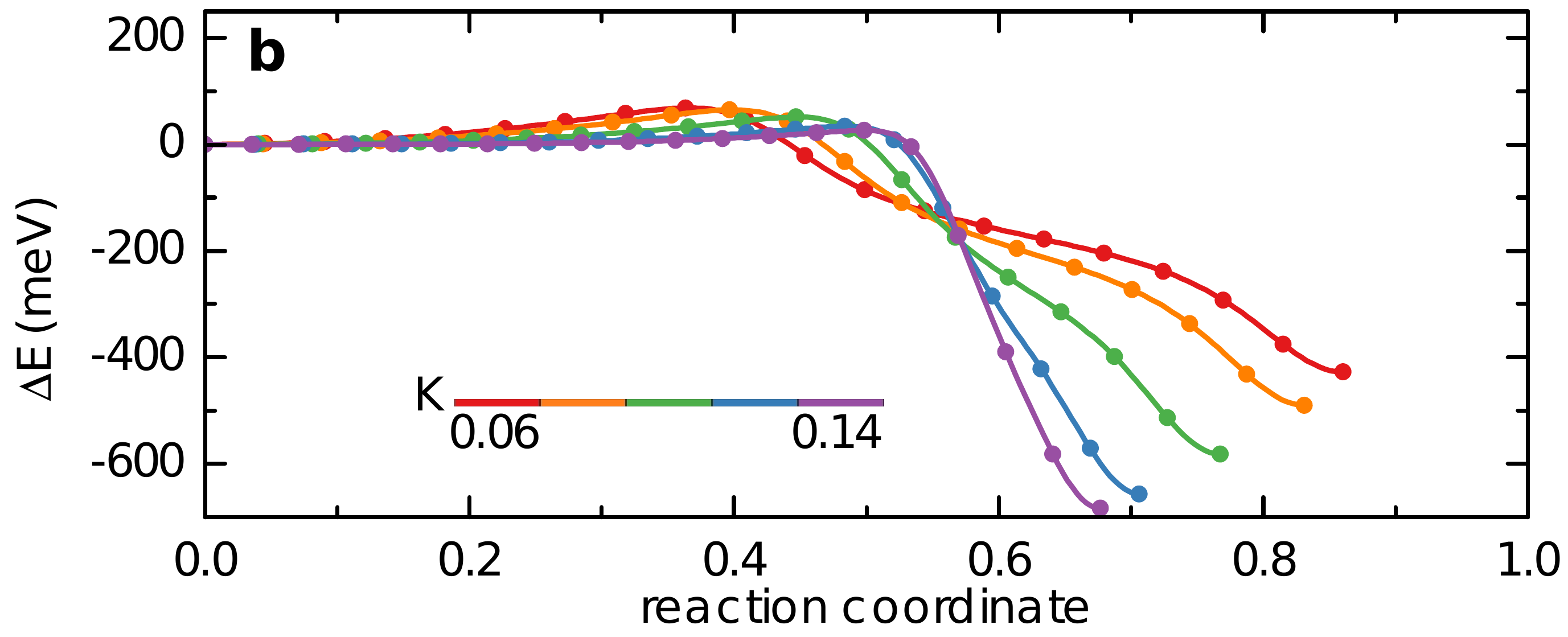}
\caption{Minimum-energy paths for (a) isotropic and (b) boundary skyrmion collapse in case of weak exchange coupling ($J=10$ meV per bond). Paths are shown for different magnetic anisotropies with respect to the FM ground state. Solid lines are cubic polynomial interpolations of the discrete images (dots).\label{fig10}}
\end{center}
\end{figure}

\subsection{Isotropic vs. boundary collapse}
First, we consider the case of an isotropic collapse in a 3ML Co film on Pt. The initial path is constructed by rotating each atomic spin from its initial orientation to the final one using Rodrigues' rotation formula~\cite{Bessarab}. Fig.~\ref{fig7}(a-e) shows that the converged path is characterized by a progressive reduction of the skyrmion diameter down to zero~\cite{Rohart,Lobanov}. Once the skyrmion core disappears [see Fig.~\ref{fig7}(e)], the central spins are flipped due to ferromagnetic ordering (shown in red) which leaves a ring of spins pointing in the plane (shown in yellow). This corresponds to a peak in the magnetization ($S_z$) profile as the skyrmion is pointing up in the center, reaches a maximum at the ring, and again points up outside the skyrmion region. Moreover, the observed behavior coincides with the magnetization profiles in Ref.~\onlinecite{Lobanov}, but was not found in Ref.~\onlinecite{Rohart} where the paths failed to converge to a minimum-energy path. 

The central spins in the last image are flipped due to ferromagnetic ordering, which leaves a ring of spins pointing in the plane. This is accompanied by a peak in the magnetization ($S_z$) profile as found in Ref.~\onlinecite{Lobanov}. Fig.~\ref{fig8}(a) shows that the corresponding energy barriers increase for lower magnetic anisotropy, surpassing $100$ meV for $J<30$ meV per bond. The monotonic increase with increasing $J$ follows the increase in skyrmion size, which requires more core spins to flip into the FM state along the path. We note however that the exact shape and maximum of the energy curve will depend on the geometry and size of the finite system. The obtained minimum energy paths [Fig.~\ref{fig9}(a)] indicate that the transition path is similar to the monolayer case~\cite{Rohart,Lobanov}. For a skyrmion with diameter of $4.7$ nm the energy barrier is $41$ meV compared to the $37$ meV found in Ref.~\onlinecite{Lobanov}. The low energy barrier can be explained by the competition between the destabilizing DMI which dilutes in monolayers away from the substrate and the thickness of the magnetic layer which increases skyrmion stability. This suggests that the energy barriers in the considered case are going to be similar to those in 1ML Co on Pt.

%reduced skyrmion stability because  DMI is diluted in monolayers away from the substrate is counteracted by an increased stability which arises from the thickness of the magnetic layer, and results in similar energy barriers to the monolayer case.

The second considered situation is the skyrmion collapse at the sample boundary, i.e. the physical end of the ferromagnetic film~\cite{CortesOrtuno}. In this case, an initial path is made by sliding a window containing the skyrmion from the center to the boundary of the sample. In the final transition path [shown in Fig.~\ref{fig7}(f-j)], the skyrmion core is distorted towards the boundary, before opening into a half circle, after which the spins gradually relax into the FM state. The saddle point in energy is found when the approaching skyrmion causes the boundary spins to cant outwards, such that their angle with the substrate (further denoted as $\theta$) is greater than $\pi/2$. The exact distance at which this happens can be described by the exchange interaction length $\sqrt{J/K}$, which is the length below which atomic exchange interactions dominate anisotropy or magnetostatic fields, and governs the width of the transition between magnetic domains~\cite{Abo}. In the considered few-monolayer ferromagnetic films, we find that N{\'e}el domain walls can have widths of around $3\sqrt{J/K}$, as was suggested in literature for the case of very thin films~\cite{Abo,Ramstock}.

\begin{figure}[t]
\begin{center}
\includegraphics[trim = 0 0 0 0, clip, width=\columnwidth]{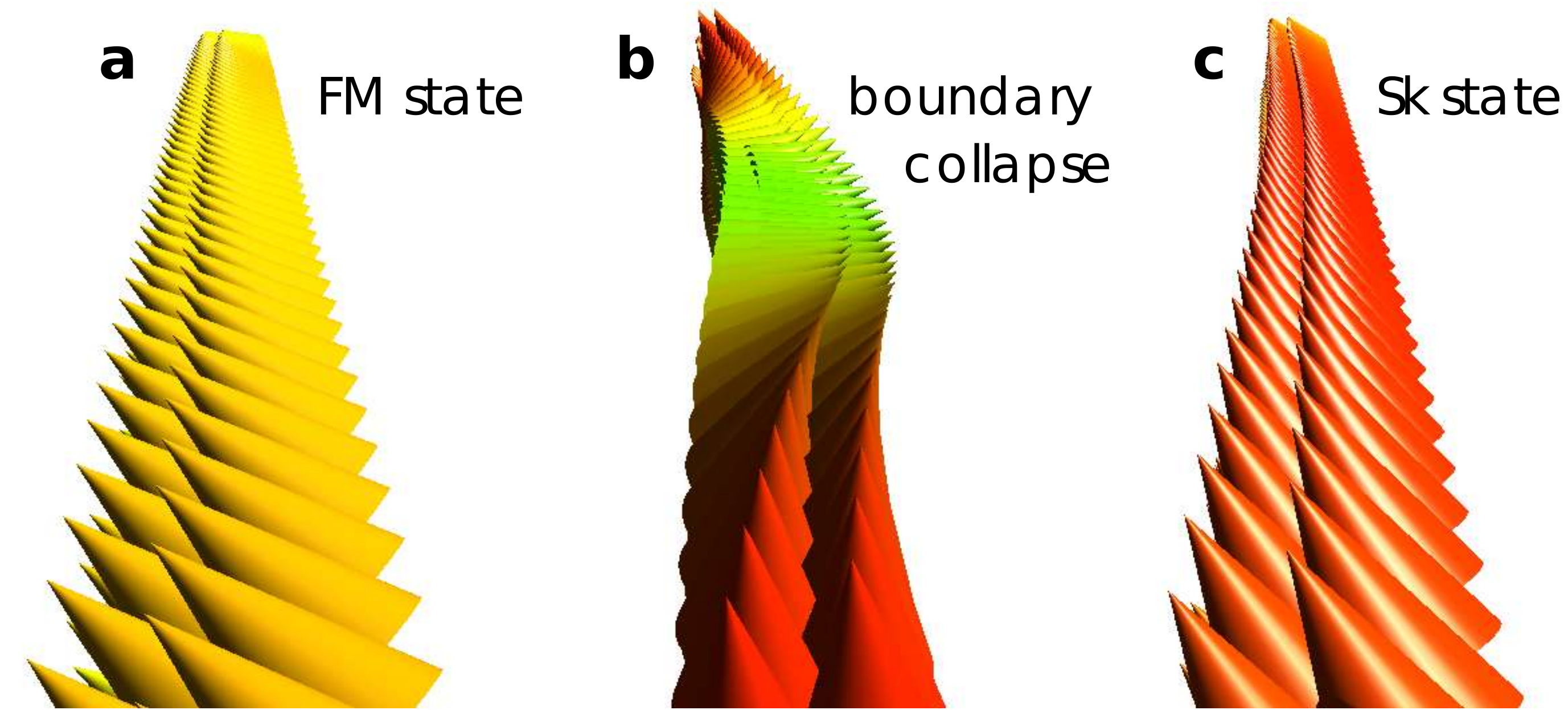}
\includegraphics[trim = 0 0 0 0, clip, width=\columnwidth]{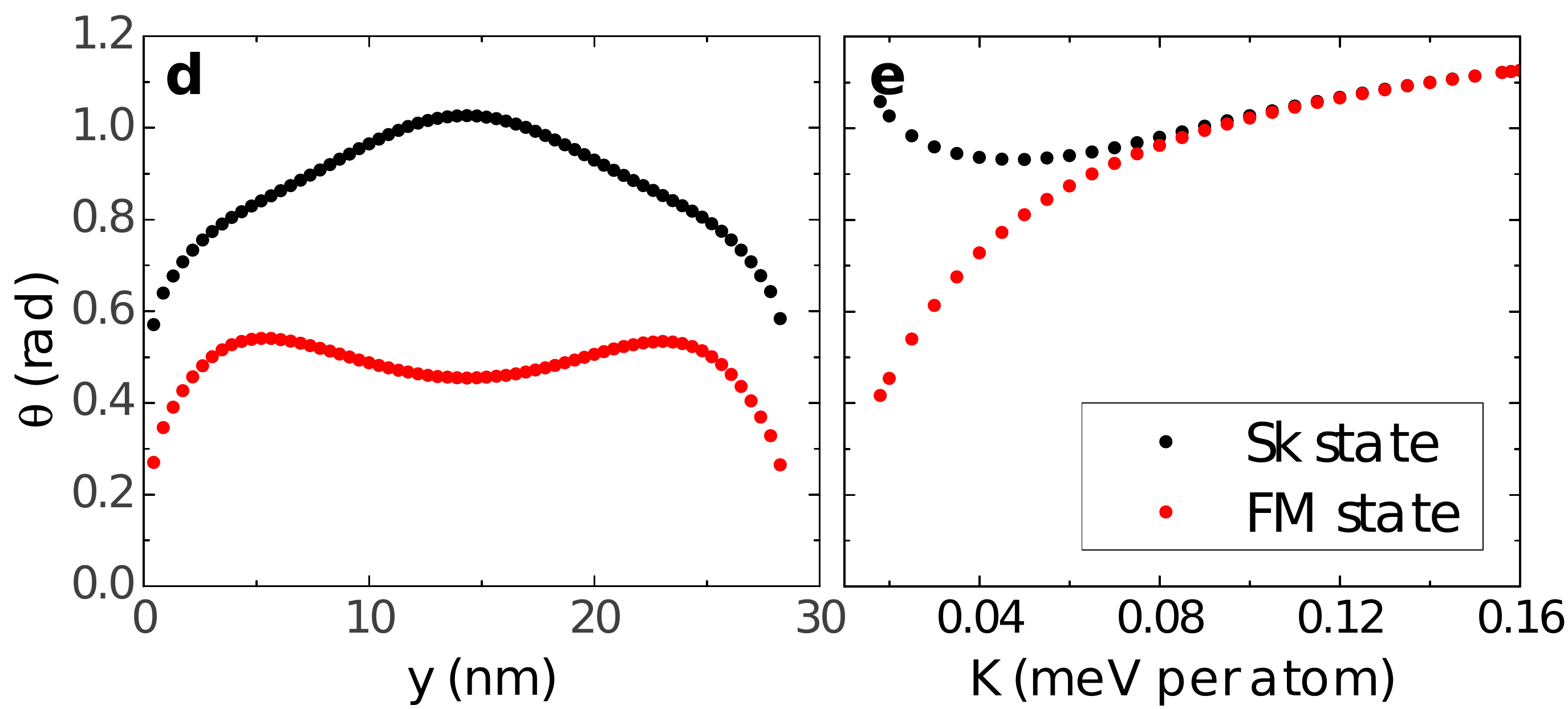}
\caption{(a) Side view of atomic spins along the right boundary of the system for the FM ground state, during boundary collapse, and the Sk state left to right respectively. The variation of the angle that boundary spins make with the substrate is shown in panel (d) as a function of position along the boundary and in panel (e) as a function of magnetic anisotropy, for the central spin on the boundary. Calculations are performed for low exchange coupling ($J=10$ meV per bond). \label{fig11}}
\end{center}
\end{figure}

\begin{figure}
\begin{center}
\includegraphics[width=\columnwidth]{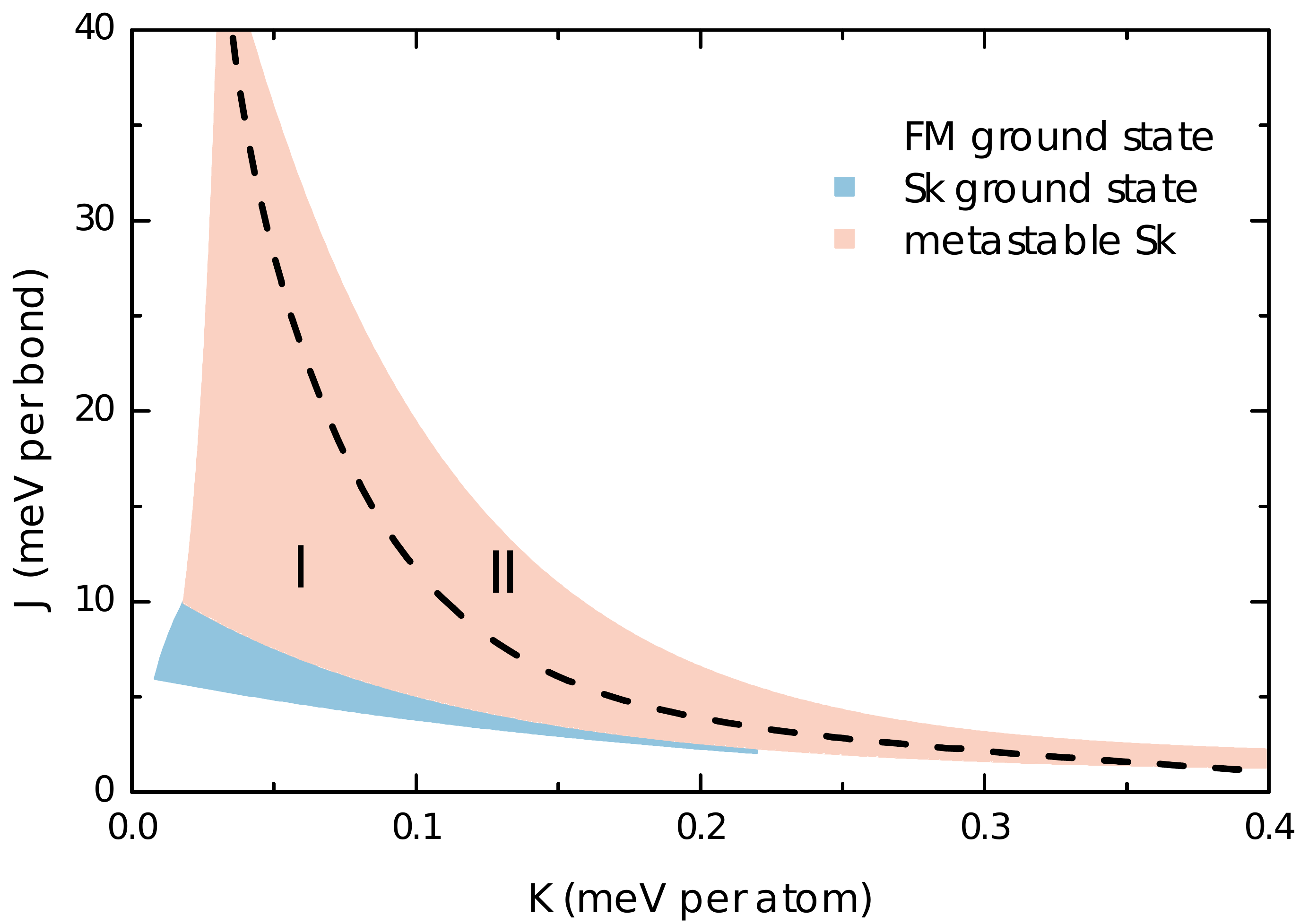}
\caption{Magnetic state diagram from Fig. \ref{fig1}, but with a dashed line separating the region where stability of the skyrmion is dictated by boundary collapse (I) from the region where the isotropic collapse dominates (II).
\label{fig1bis}}
\end{center}
\end{figure}

\begin{figure*}[t]
\begin{center}
\includegraphics[trim = 0 0 0 0, clip, width=\textwidth]{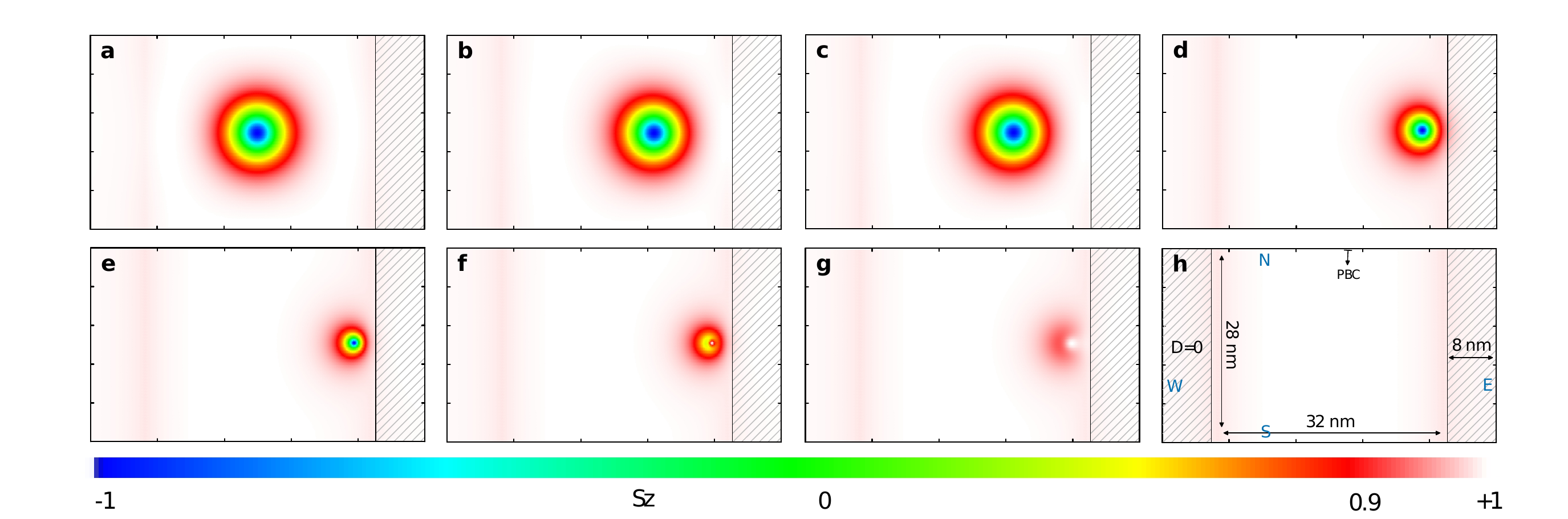}
\caption{Top view of spatial magnetization distribution [$S_z(x,y)$] in a 3ML Co on Pt, for skyrmion collapse at the interface with boundary region (of width $8$ nm, see inset) where DMI vanishes. The parameters used are $J=30$ meV per bond for exchange and $K=0.050$ meV per atom for anisotropy. Individual images correspond to the initial Sk state (a), path towards the saddle point (b-d), and subsequent stages (e,g) towards the FM ground state (h).
\label{fig12}}
\end{center}
\end{figure*}

Fig.~\ref{fig8}(b) shows that the energy barriers in a boundary collapse are significantly different than those found for the isotropic collapse [Fig.~\ref{fig8}(a)], presenting a maximum for intermediate anisotropy instead of the monotonic decrease with anisotropy. Further investigation shows that the observed maxima in barriers is in fact the boundary effect due the confined geometry of the sample. Due to open boundary conditions, one finds a spontaneous inward canting of the boundary spins~\cite{Rohart1} even for the FM ground state [shown in Fig.~\ref{fig11}(a)]. However, the presence of a large skyrmion (for example for weak exchange interaction) can affect the boundary spins in such a way that the tilt angle increases [i.e. $\theta$ approaches $\pi/2$, see Fig.~\ref{fig11}(b)] and the energy required to collapse the skyrmion (or equivalently, to surpass the canting of boundary spins) is reduced. For given size of the system, the skyrmion size for which the latter canting occurs will depend on $l_{ex}$. For example, the peak energy barrier occurs at skyrmion diameter of $12$ nm for low exchange interactions ($J<15$ meV per bond), whereas for high exchange couplings skyrmions of only a few nanometers in size can already affect the boundary spins in our samples.

Minimum energy paths for skyrmion collapse at the boundary are shown in Figs.~\ref{fig9}(b) and~\ref{fig10}(b), for strong and weak exchange respectively. For strong $J$ the paths have a similar behavior to the paths found for the isotropic collapse. % a slow increase in energy up to the saddle point, after which they gradually descend to the ground state energy, and are thus not significantly different from the paths found for the isotropic collapse.
On the other hand, paths for weak $J$ are characterized by two stages corresponding to the (i) skyrmion reaching the boundary, and (ii) skyrmion fully exiting on the boundary. This step is found when the exchange length is smaller than the skyrmion diameter, thus becomes less pronounced for stronger exchange couplings or anisotropies (where $l_{ex}$ exceeds the skyrmion size).

The calculated energy barriers for the isolated and boundary collapse, presented in Fig.~\ref{fig8}(a,b) enable us to determine the preferential collapse mechanism of a skyrmion in a 3ML Co on Pt, in the presence of e.g. thermal fluctuations. The found preferential collapse mechanisms, superimposed on the parametric range of stability for a skyrmion (same as in Fig. \ref{fig1}) are indicated in Fig. \ref{fig1bis}. For skyrmions near the upper bound of their stability area in the magnetic state diagram of Fig.~\ref{fig1bis}, the isotropic collapse has a smaller energy barrier compared to the collapse at the boundary. Therefore we expect the isotropic collapse to be deterministic for the stability of skyrmions in that parametric range and not the boundary collapse as suggested in Ref.~\onlinecite{CortesOrtuno}. The energy barriers for the isotropic collapse can be increased by lowering the anisotropy, i.e. increasing the skyrmion size, which would however strongly decrease the barrier for the boundary collapse. We conclude that for small skyrmions [region I in Fig.~\ref{fig1bis}] the isotropic collapse limits their stability, whereas for large skyrmions [region II in Fig.~\ref{fig1bis}] their stability is dictated by the boundary collapse.

Finally we briefly discuss the comparison to the minimum energy paths for skyrmions in case of homogeneous DMI for all three Co monolayers, given by black lines in Fig.~\ref{fig9}. The collapse behavior for isotropic and boundary transitions is seemingly similar to those obtained in the monolayer-resolved case for 3ML Co on Pt. This is expected since we found earlier that the skyrmion persists through all three monolayers, with nearly equal size and energy as for monolayer resolved DMI. However, the energy barriers are significantly different. The isotropic collapse has an energy barrier of $336$ meV for homogeneous DMI compared to the maximum of $74$ meV found for monolayer-resolved DMI, whereas the boundary collapse occurs at $172$ meV compared to $22$ meV respectively. This is another manifestation of the fact that the increased cumulative DMI of $4.5$ meV per bond ($1.5$ meV for each monolayer) induces more stable Sk states, since in the layer-resolved case DMI is diluted in monolayers away from the interface with Pt.

\subsection{Collapse at lateral DM interfaces\label{dminterface}}
It is clear that skyrmions have poor thermal stability in finite systems due to proximity of the boundaries, which complicates their practical applications in e.g. racetrack memories. For skyrmions to be usable in desirable situations, such as near room temperature and in nanostructured circuits, one needs to increase the energy barrier for the collapse of the Sk state. One recently reported way to achieve this is by lateral heterostructuring of the DMI~\cite{Mulkers1}, so that the confining system is not determined by the finite size of the ferromagnetic film but rather by vanishing DMI (due to e.g. patterned heavy-metal layer(s)). In what follows, we consider the same system (3ML Co on Pt) with monolayer resolved DMI inside, surrounded by an outer region with no DMI, where then an open boundary is placed. We thus study the stability of a skyrmion towards collapse at the interface between regions with different DMI. In Fig.~\ref{fig12} we illustrate the considered system, where we extended the previous sample by a surrounding band of width 8 nm where DMI vanishes (so system now contains a total of $75660$ spins). We considered several initial paths: sliding a window containing the skyrmion from the center to the boundary, relaxing each image by fixing the core spin, and  using the boundary transition as input. As shown in Fig.~\ref{fig12}(a-g), the converged path involves a skyrmion translation towards the interface and a rotation of the spins into the FM state. The rotation is similar to that of the isotropic collapse and occurs right before the interface. The fact that we do not observe an escape through the interface like in the boundary case can be explained by the inclusion of the magnetic layer instead of a vacuum. In particular, the ferromagnetic ordering between spins near the interface, which point in opposite direction of the skyrmion core, provides a strong repulsive interaction against the skyrmion core.
\begin{figure}[t]
\begin{center}
\includegraphics[width=\columnwidth]{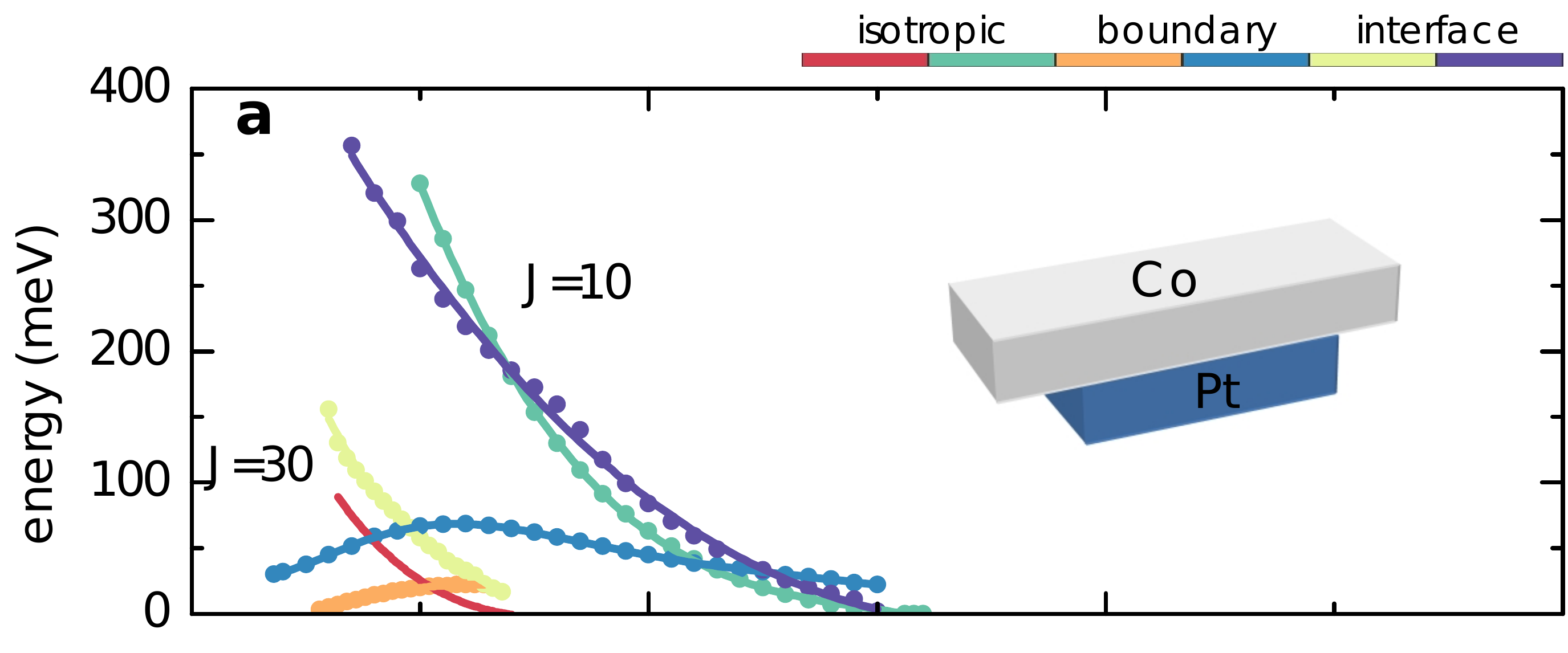}
\includegraphics[width=\columnwidth]{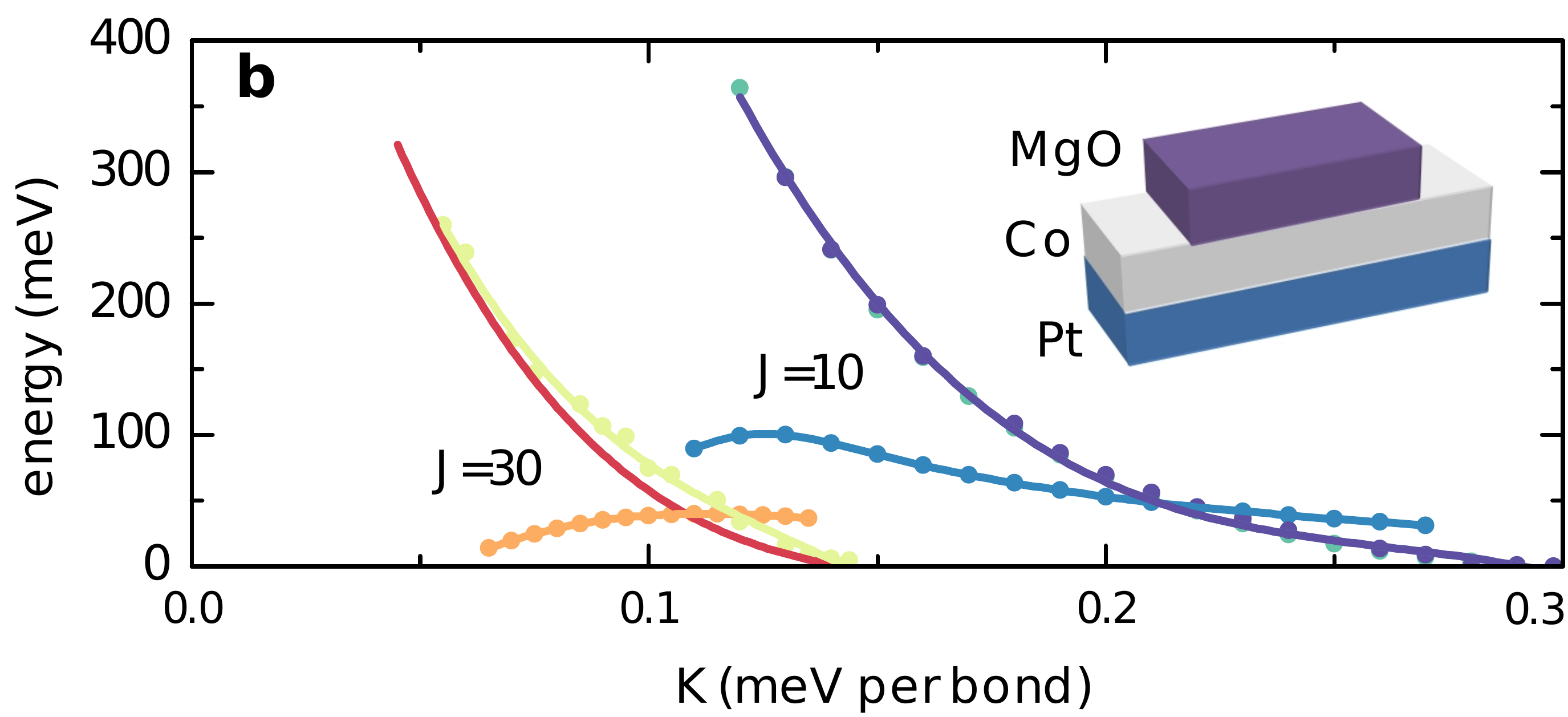}
\caption{(a) Energy barriers for the collapse of a skyrmion in the sample shown in Fig. \ref{fig12}(h) (3ML Co on Pt, with monolayer resolved DMI), at a lateral (East/West) interface where DMI vanishes, at North/South open boundary of the film, and for the isotropic collapse, as a function of magnetic anisotropy. Each point corresponds to a unique energy barrier retrieved from the maximum (saddle point) in a minimum energy path. (b) Same as (a), but for the Pt-3ML Co-MgO sandwich, with monolayer resolved DMI, and removed MgO on lateral sides of the sample (thus lower DMI, particularly in the top Co monolayer). In either sample, stability of the skyrmion is dictated by boundary collapse at low anisotropy and isotropic collapse otherwise. The barrier for collapse at DMI interface greatly exceeds the one for boundary collapse at all $(J,K)$.
\label{fig13}}
\end{center}
\end{figure}

Fig.~\ref{fig13}(a) reveals that the energy barrier for collapse of the skyrmion at the interface where DMI vanishes (East/West side, see sample in Fig. \ref{fig12}), is significantly larger than the barrier for a transition at the open boundary. The barrier for collapse at the DMI interface is actually rather similar to the one for the isotropic collapse, since the spin-flipping sequence along the minimal energy path is essentially the same in those two cases. This implies that the the favorable collapse mechanisms in the $(J,K)$ diagram for the considered system with supressed DMI on the two sides will be nearly the same as for the sample without any DMI interface. However, our results also mean that if we were to completely surround the inner region by a region with low DMI, or elongate the sample vertically, the boundary collapse will be suppressed and the barrier for skyrmion collapse for all $(J,K)$ combinations will be increased. This is very beneficial for future design of e.g. skyrmion racetracks, by having an extended Co film on a track defined by the pre-patterned Pt substrate.

Arguably, it is more convenient to use the extended Pt substrate and Co film on it, but then pattern the MgO capping layer to form a racetrack. In Fig.~\ref{fig13}(b) we show the barriers for different types of skyrmion collapse for this case, for the sample size identical to one of Fig.~\ref{fig13}(a). The interior of the sample now has MgO on top, with corresponding layer-resolved DMI (high in the top Co monolayer, see Fig. \ref{fig5}), while the outer region is bare 3ML Co on Pt - with significantly lower DMI in the top Co monolayer. In this case, the stability of the skyrmion is improved as energy barriers for the skyrmion collapse are higher [shown in Fig. \ref{fig13}(b)], but the relationships between different mechanisms for collapse remain similar to the previously considered sample [in Fig.~\ref{fig13}(a)]. 

\begin{figure}
\begin{center}
\includegraphics[width=\columnwidth]{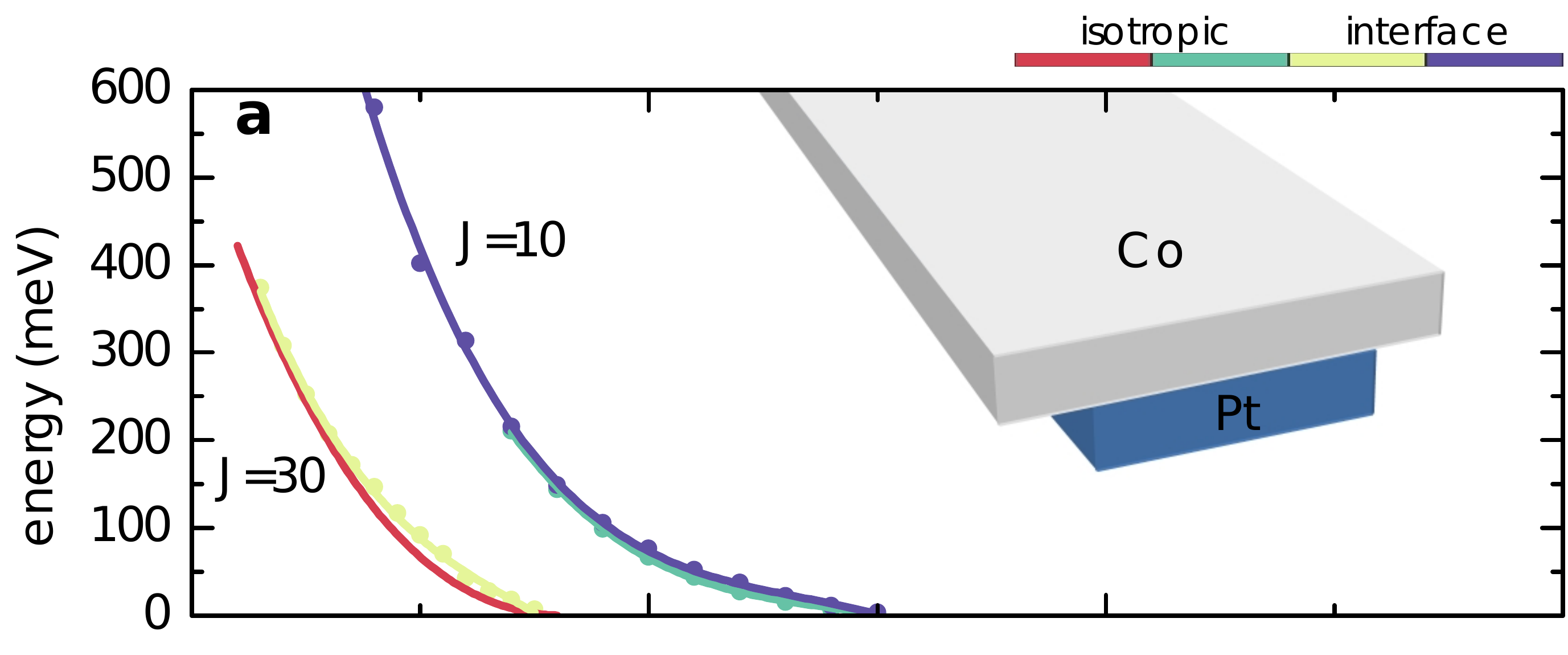}
\includegraphics[width=\columnwidth]{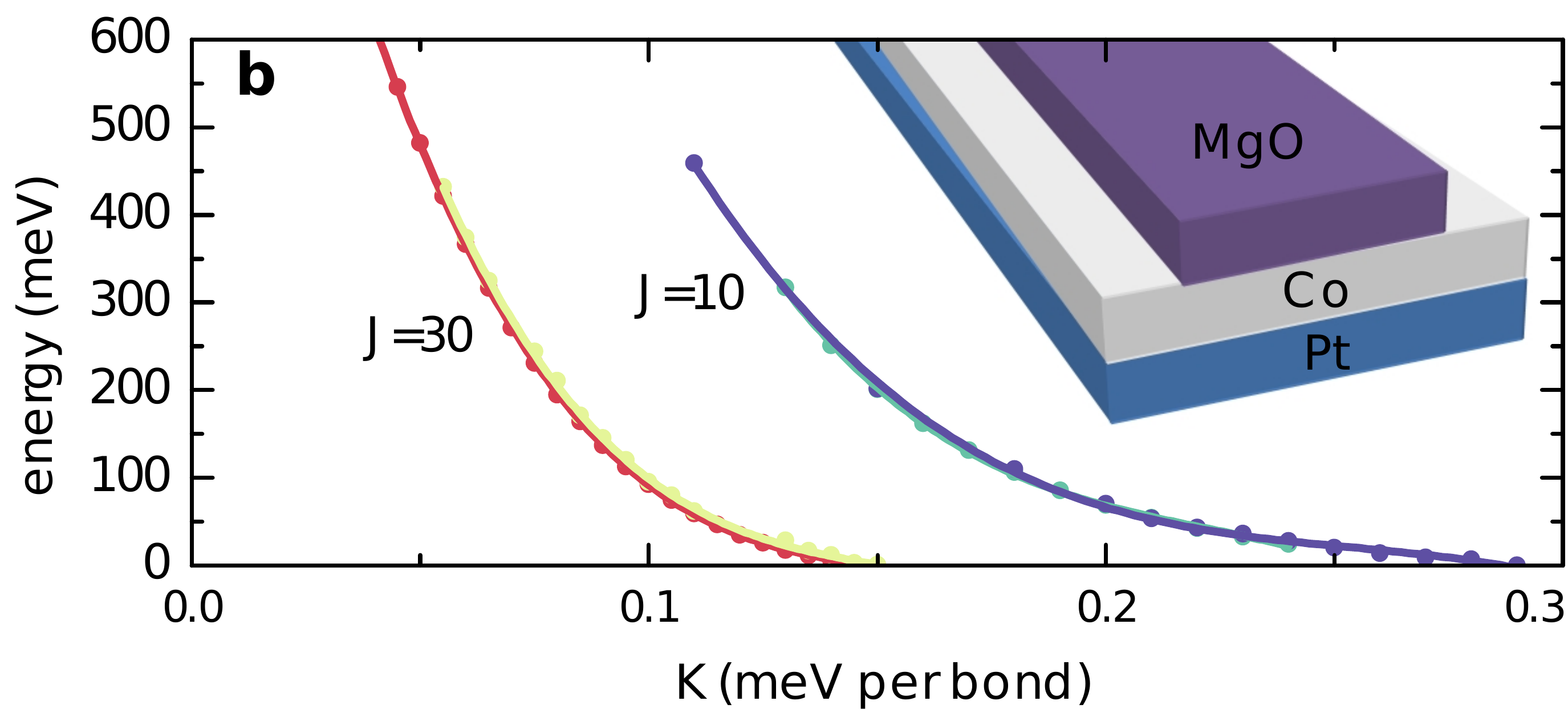}
\includegraphics[width=\columnwidth]{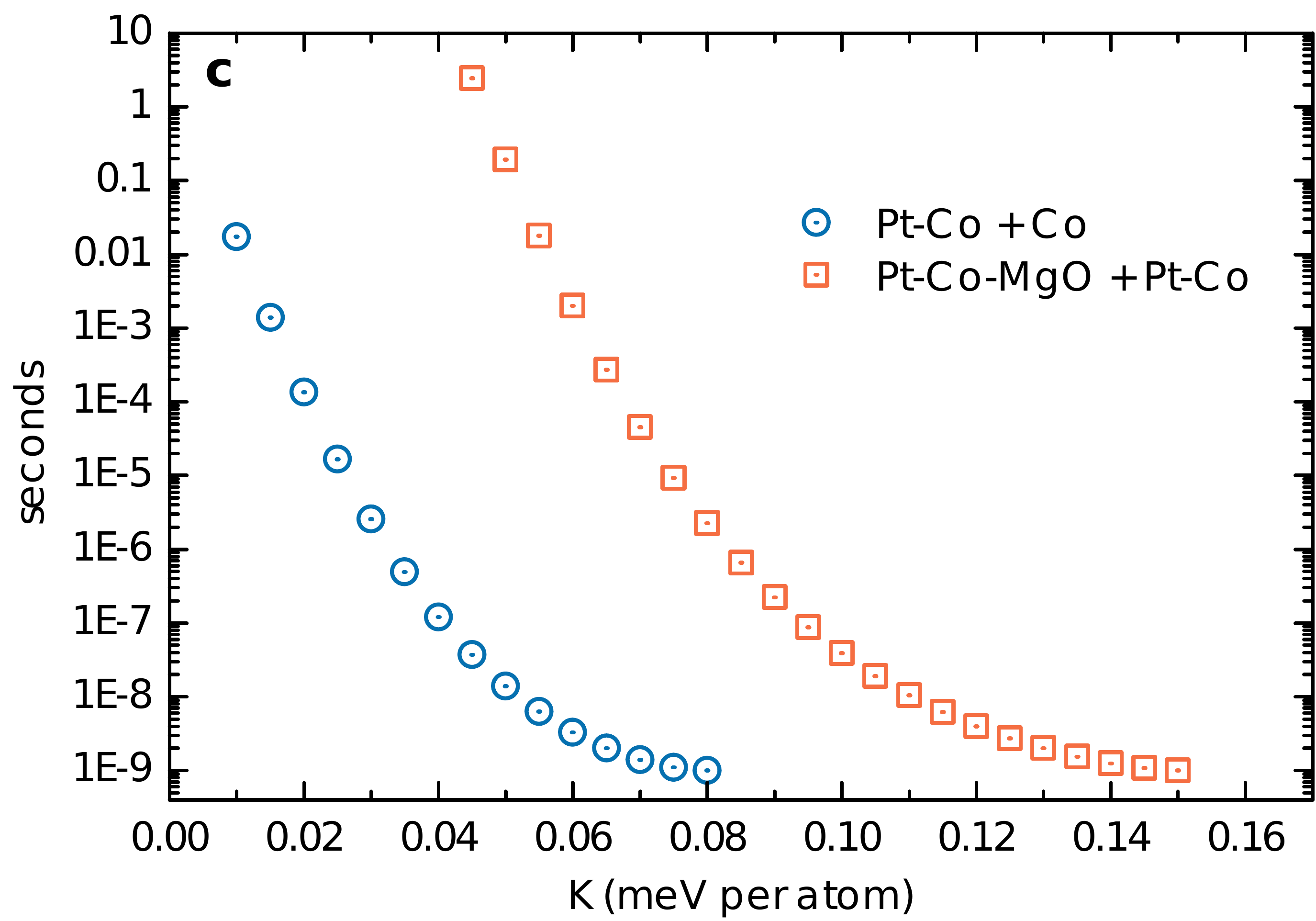}
\caption{(a,b) Same as Fig. \ref{fig13}(a,b) respectively, but for periodic boundary conditions used at North/South edges of the sample, so that effectively an elongated track with laterally suppressed DMI is considered. (c) The lifespan of an isolated skyrmion [given by Eq. (\ref{eqt})] calculated for the most favorable mechanism of collapse in (a,b), for standard exchange coupling for Co ($J=30$ meV per bond), as a function of anisotropy, at room temperature.
\label{fig14}}
\end{center}
\end{figure}

Therefore, as an optimized design of the skyrmion racetrack at room temperature we discuss the case of the same geometry of the sample as in Fig. \ref{fig12}, but with periodic boundary conditions on North/South edges. The behavior of the barriers for skyrmion collapse, now that there is no open boundary, are shown in Fig. \ref{fig14}(a,b), for the materials considered in Fig. \ref{fig13}(a,b) respectively. We indeed confirm that the barriers for collapse are significantly higher, particularly in the low-anisotropy regime where boundary collapse usually dominates.

Next, from these optimized energy barriers in Fig. \ref{fig14}(a,b), it is interesting to quantify the lifetime of skyrmions using the Arrhenius law
\begin{align}
\tau=\tau_0\exp\left(\frac{\Delta E}{k_BT}\right),
\label{eqt}
\end{align}
where $\tau_0$ is related to the attempt frequency $f_0=1/\tau_0$, $\Delta E$ is the energy barrier, $k_B$ is the Boltzmann constant, and $T$ is the temperature. Since the attempt frequencies are hard to obtain and can vary in magnitude, we assume $f_0=10^9$ Hz as suggested in literature\cite{Rohart,CortesOrtuno}. We focus on the skyrmion lifetime for the considered systems (see Figs.~\ref{fig13} and Fig.~\ref{fig14}) and scan the lowest energy barrier for skyrmion collapse as the effective anisotropy is varied. Fig.~\ref{fig14}(c) shows that the lifespan of the skyrmion at room temperature spans a wide range of magnitudes, from the order of seconds to just a few nanoseconds, but we can determine the relevant values based on parameters retrieved from experimental works~\cite{Boulle,Moreau}. The exchange constant for the Co layer is roughly $30$ meV per bond. The effective anisotropy varies between $0.019$ meV per atom for Pt-Co-MgO sandwich~\cite{Boulle} and $0.015$ meV per atom for Pt-Co-Ir multilayers~\cite{Moreau} (conversion relations between micromagnetic and atomistic parameters can be found in Ref.~\onlinecite{CortesOrtuno}). For such parameter values the skyrmion lifetime can reach few seconds, based on isotropic or interfacial collapse discussed in Fig.~\ref{fig14}. The boundary escape is not a realizable collapse mechanism in the considered system, which has interfaces with regions of no DMI in East/West direction the skyrmion cannot surpass (see previous discussions), and is taken periodic in North/South direction so there is no boundary there. However, for systems with open boundary conditions in the North/South direction, the boundary collapse has lifetimes that can reach maximum of few nanoseconds, as seen from the energy barriers in Fig.~\ref{fig13}.

A proper analysis of skyrmion lifetimes is difficult as the exact collapse mechanism which the skyrmion will undertake depends on its distance from the boundary or interface. For example, a skyrmion positioned at the center of a very large sample will naturally be limited by the isotropic collapse, whereas a skyrmion near the edge might escape at the boundary instead. Ref.~\onlinecite{Woo} has indeed observed long-lived skyrmions near room-temperatures, albeit on racetracks that are $20$ times wider than the skyrmion diameter, where escape at the boundary is very unlikely to occur. In nanometer-wide racetracks which are suitable for ultradense magnetic storage devices, however, the diameter of the skyrmion is only two to five times smaller than the width of the racetrack, and we can therefore expect that collapse at the boundary is inevitable and will dictate skyrmion stability. We therefore believe that our analysis on skyrmion lifetimes matches the behavior of room-temerature skyrmions observed in experiments. These results indicate that lifetime of a skyrmion can reach seconds in the considered samples, compared to maximum few nanoseconds in same geometries with open boundary. This confirms again that the DM interface as proposed here could be a very convenient solution to the weak stability of skyrmions in finite systems.

%However, we can determine which parameters are relevant based on recent experimental works~\cite{Boulle,Moreau}. In particular, we know that the exchange interaction is strong ($J=30$ meV per bond), while the effective anisotropy varies between $K=0.019$ meV per atom for the Pt-Co-MgO system~\cite{Boulle} and $K=0.015$ meV per atom for the Pt-Co-Ir mulitlayer~\cite{Moreau}. The relations used to convert the micromagnetic parameters into atomistic values can be found in Ref.~\onlinecite{CortesOrtuno}. These results indicate that lifetime of a skyrmion can reach seconds in the considered samples, compared to maximum few nanoseconds in same geometries with open boundary. This confirms again that the DM interface as proposed here could be a very convenient solution to the weak stability of skyrmions in finite systems.

\section{Conclusions\label{concl}}
In summary, we have studied the stability of isolated skyrmions in ultrathin ferromagnetic films with interfacially induced DMI - specifically three monolayers of Co on Pt, with or without a capping spin-orbit layer (MgO, Ir, Pt, or similar). The consideration of several magnetic monolayers, along with monolayer-resolved DMI retrieved from first principles~\cite{Yang} gives rise to more realistic, but less stable Sk states than previously considered in literature. The obtained skyrmions have different core structure and consist of fewer core spins and in different energy profiles per atom, strongly dependent on the realized DMI profile. We discussed three possible mechanisms for skyrmion collapse in these realistic systems, via isotropic rotation of the spins, collapse at the boundary of a laterally finite film, or collapse at the lateral interface where DMI changes. We find that the stability of skyrmions smaller than few nm is limited by the isotropic collapse, whereas larger ones are more likely to collapse at the boundary. The confined geometry in finite systems causes poor thermal stability of the skyrmion~\cite{CortesOrtuno}, as the skyrmion sees low energy barrier toward collapse at the boundary. The correspondingly short skyrmion lifetime severely limits their applicability, particularly at room temperature. Instead, we show that it is most favorable to replace the boundary by an interface where DMI is lowered~\cite{Mulkers1}, by e.g. having an extended Co film on Pt, but finite capping layer of MgO on top \cite{Boulle}. In such a case, the barrier for skyrmion collapse at the interface where DMI changes is far larger than it used to be at the open boundary, and lifetime of the skyrmion at room temperatures extends from few nanoseconds to few seconds, making skyrmions much more favorable for future data processing devices.

\begin{acknowledgments}
This work was supported by the Research Foundation - Flanders (FWO-Vlaanderen) and Brazilian agencies CNPq and FACEPE.
\end{acknowledgments}

\input{mulsk.bbl}
\end{document}

%% file: mulsk.bbl
\providecommand{\noopsort}[1]{}\providecommand{\singleletter}[1]{#1}%